\documentclass[a4paper, superscriptaddress, amsfonts, amssymb, amsmath, reprint, showkeys, nofootinbib, twoside]{revtex4-1}
\usepackage[english]{babel}
\usepackage[utf8]{inputenc}
\usepackage[colorinlistoftodos, color=green!40, prependcaption]{todonotes}
\usepackage{amsthm}
\usepackage{mathtools}
\usepackage{physics}
\usepackage{xcolor}
\usepackage{graphicx}
\usepackage[left=23mm,right=13mm,top=35mm,columnsep=15pt]{geometry} 
\usepackage{subfigure}
\usepackage{multirow}
\usepackage{adjustbox}
\usepackage{placeins}
\usepackage[T1]{fontenc}
\usepackage{lipsum}
\usepackage{csquotes}
\usepackage[pdftex, pdftitle={Article}, pdfauthor={Author}]{hyperref} % For hyperlinks in the PDF
\begin{document}
\title{Using Deep Learning to Localize Gravitational Wave Sources}

\author{Chayan Chatterjee}
\email{chayan.chatterjee@uwa.edu.au}
\author{Linqing Wen}%
 \email{linqing.wen@uwa.edu.au}
\affiliation{%
 Department of Physics, The University of Western Australia,\\ 35 Stirling Hwy, Crawley, Western Australia 6009, Australia
}
 
\author{Kevin Vinsen}
\email{kevin.vinsen@icrar.org}
\affiliation{
 International Centre for Radio Astronomy Research, The University of Western Australia\\ M468, 35 Stirling Hwy, Crawley, WA, Australia
% %This line break forced% with \\
}%
 
 \author{Manoj Kovalam}
 \email{manoj.kovalam@research.uwa.edu.au}
\affiliation{%
Department of Physics, The University of Western Australia,\\ 35 Stirling Hwy, Crawley, Western Australia 6009, Australia
}%

\author{Amitava Datta}
\email{amitava.datta@uwa.edu.au}
\affiliation{%
 Department of Computer Science and Software Engineering, The University of Western Australia\\
 35 Stirling Hwy, Crawley, Western Australia 6009, Australia
 }%

\date{\today} % Leave empty to omit a date

\begin{abstract}
Deep Learning algorithms, in particular neural networks have been steadily gaining popularity among the gravitational wave community for the last few years. The reliability and accuracy of Deep Learning approaches in gravitational wave detection, parameter estimation and glitch classification have already been proved and verified by several groups in recent years. In this paper, we report on the construction of a deep Artificial Neural Network (ANN) to localize simulated gravitational wave signals in the sky with high accuracy. We have modelled the sky as a sphere and have considered cases where the sphere is divided into 18, 50, 128, 1024, 2048 and 4096 sectors. The sky direction of the gravitational wave source is estimated by classifying the signal into one of these sectors based on it's right ascension and declination values for each of these cases. In order to do this, we have injected simulated binary black hole gravitational wave signals of component masses sampled uniformly between 30-80 $M\textsubscript{\(\odot\)}$ into Gaussian noise and used the whitened strain values to obtain the input features for training our ANN. We input features such as the delays in arrival times, phase differences and amplitude ratios at each of the three detectors Hanford, Livingston and Virgo, from the raw time-domain strain values as well as from analytical versions of these signals, obtained through Hilbert transformation. We show that our model is able to classify gravitational wave samples, not used in the training process, into their correct sectors with very high accuracy ($>$ 90\%) for coarse angular resolution using 18, 50 and 128 sectors. We also test our localization on test samples with injection parameters of the published LIGO binary black hole merger events GW150914, GW170818 and GW170823 for 1024, 2048 and 4096 sectors and compare the result with that from BAYESTAR and Parameter Estimation (PE). In addition, we report that the time taken by our model to localize one GW signal is around 0.018 secs on 14 Intel Xeon CPU cores.   
\end{abstract}


\maketitle

%\input{sections/section01.tex}  %I believe leaving the sections in separate files is more organized, change it if you desire 
%\input{sections/section02.tex}
%\input{sections/section03.tex}
%\input{sections/acknowledgements.tex}

%\bibliography{ms}

\section{\label{sec:level1}INTRODUCTION}

The Laser Interferometer Gravitational Wave Observatory (LIGO) \cite{PhysRevLett.116.131103,LIGO} and its European counterpart, Virgo \cite{virgo}, have made, at the time of writing, more than twenty reported detections of gravitational waves (GW) since their inception \cite{bp1,bp2,bp3,bp4,bp5,bp6,bp7}. 
These GW signals, produced by the coalescence of binary black hole \cite{bp1,bp2,bp3,bp4} and binary neutron star systems (BNS) \cite{bp5,bp6,bp7}, have made Gravitational Wave Astronomy a reality and it is now a major component of multi-messenger astronomy, alongside neutrino and cosmic ray~\cite{astro_1,astro_2,astro_3} observations.
With more detectors and better sensitivities, new GW sources will be detected on a regular basis and faster follow-up of electromagnetic (EM) counterparts of these mergers will provide detailed information about the origin and composition of these compact objects. \cite{em_1,em_2,em_3,em_4,em_5,em_6}. 

LIGO and Virgo began their third observation run (O3) in April, 2019 and have already detected several new GW signals. The technique used by search pipelines to identify and characterize GW signals from noise is matched filtering, which uses a bank of template waveforms of different component masses and spins \cite{mf1,mf2,mf3,mf4,mf5,mf6,mf7,mf8}. These waveform models cover the inspiral, merger and ringdown phases of a compact binary coalescence and are generated by combining post-Newtonian theory \cite{pn1,pn2,pn3,pn4}, the effective-one-body formalism \cite{eob} and numerical relativity simulations \cite{NumRel}. While this approach has been very successful, the algorithms used are computationally expensive due to the fact that they probe a large parameter space with increasingly longer-duration waveforms, as the low-frequency sensitivity of the interferometers improves. Also, future observation runs will involve more detectors running simultaneously over longer periods of time, thereby greatly increasing the bulk of data to be analysed by signal processing techniques. This also makes it necessary to accelerate parameter estimation algorithms which normally takes several hours to process \cite{pe_fast}.

Machine Learning algorithms, like deep learning \cite{deeplearning1,deeplearning2,deeplearning4,deeplearning5,deeplearning6}, can be used to address these issues since the only computationally intensive step is the one-time training phase that happens prior to the analysis of actual data. This enables low latency detection, as well as fast parameter estimation, making this approach possibly orders of magnitude faster than other techniques. 

The important parameters used for localizing the GW sources are its Right Ascension (RA) and Declination (Dec) values which are its coordinates in the sky. It is necessary to obtain RA and Decs with minimum uncertainty so that EM telescopes can be pointed in the precise direction, in real time so that extremely directional and transient counterparts like kilonova and short gamma ray bursts following binary neutron star mergers can be observed. In O3 for example, for BNS events, the median sky localisation accuracy in terms of the 90\% credible area is 120-180 deg$^2$. 12-21\% of BNS mergers are expected to be  localised to less than 20 deg$^2$ \cite{O3localization}.

\begin{figure*}
\includegraphics[width=10cm]{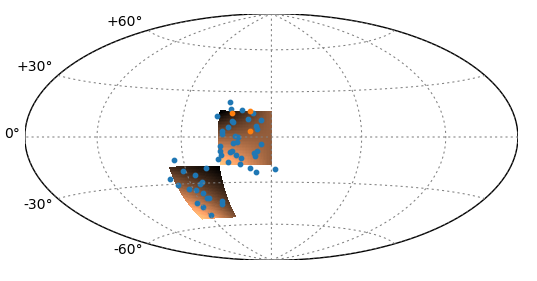}% Here is how to import EPS art
\caption{\label{fig:2} Accuracy of localization by our ANN model, described in Section VIII. The dots here are the exact sky directions of a few of our injected GW test samples. Orange dots (three of them in this case) are injections that belong to the colored sectors but were classified into immediate neighbouring sectors. Sky blue dots represent injections classified into the two sectors. Sky directions of injections lying outside the two sectors can be further correctly classified by using a multi-labelling scheme. } 
\end{figure*}

In this paper we present a new method of localizing GW sources using Artificial Neural Network (ANN) \cite{ANN}, a very popular Deep Learning algorithm. We have modelled the sky as a sphere divided into many sectors and used our network to classify simulated GW signals injected in Gaussian noise into one of the sectors of the sphere. As the number of sectors is increased, the area of the individual sectors decreases, thereby potentially helping improve the angular resolution. This, coupled with the low computational costs and high speed of machine learning algorithms makes our approach a viable option for future localization endeavours.
Figure~\ref{fig:2} shows the localization accuracy of our model for 50 sectors (For details, see Section VIII).

Section \ref{Section 2} gives a brief overview of the applications and successes of machine learning and in particular, deep learning in GW research. Section \ref{Section 3} describes the method of triangulation of three detectors which helps constrain the sky direction of the source of GW signals. Section \ref{Section 4} discusses briefly the method of ANNs and their applications. Section \ref{Section 5} describes the sample generation code used to create the GW signals and noise samples used for training and testing. Section \ref{Section 6} describes in detail the architecture of our ANN model, the rationale behind our chosen input features and the signal processing techniques used to extract those features. In Section \ref{Section 7} we describe the details of the experiments carried out for GW signals with and without noise and discuss the results and accuracy obtained for different number of sectors and SNR ranges. In Section \ref{Section 8} we evaluate the performance of our classification algorithm and describe the metrics used for the evaluation. We also highlight ways to improve the accuracy in our future work. In Section \ref{Section 9} we use our model to test samples simulated with published BH parameters for events GW150914, GW170818 and GW170823 injected into Gaussian noise and with advanced LIGO PSD. Lastly, we conclude in Section \ref{Section 10} with discussions and plans for future work.

\section{Machine Learning in Gravitational Wave Research }\label{Section 2}

Machine Learning algorithms, in particular deep learning, have already found many applications in GW research and is quickly gaining popularity as a viable alternative to the more computationally intensive, template based matched filtering algorithm \cite{ComputationallyIntensive}. The reliability of deep learning as a powerful classification and regression tool is firmly established by its varied applications in myriad sectors. Apart from GW, deep learning has been applied in image processing and classification \cite{image1,image2}, self-driving cars, medical diagnostics etc. 

In GW research, deep learning algorithms, in particular convolutional neural networks (CNN) \cite{deeplearning5} have been used to match the accuracy of matched filtering based search pipelines in signal detection and parameter estimation \cite{detection1,detection2,detection4,detection5,detection6,detection7,detection8}. It has also been extensively applied in glitch classification \cite{glitch1,glitch2,glitch3,glitch4,glitch5, glitch6}, signal denoising \cite{denoising1,denoising2,denoising3,denoising4} and has been used with numerical relativity simulations to detect higher order multipole waveforms for eccentric binary mergers \cite{multipole}. The preferred machine learning algorithms of choice are mostly different types of neural networks, with the most popular being CNNs, used by several groups for detection, parameter estimation and glitch classification. Autoencoders, another popular deep learning algorithm has been used in denoising GW signals from both parametric and real LIGO noise, with very promising results. 

%%Other popular algorithms of choice include Artificial Neural Networks (ANN), Recurrent Neural Networks (RNN), in particular Long Short Term Memory (LSTM), Random Forests (RF), Support Vector Machines (SVM) etc. 

\section{Localization of Gravitational Waves}\label{Section 3}

As mentioned in section \ref{sec:level1}, it is of pivotal importance to have accurate estimates of the sky direction of GW signals in the sky the EM telescopes have small field of view and need an accurate GW localization. It is also possible to have EM counterparts from BBH mergers \cite{EMBlackHole1,EMBlackHole2,EMBlackHole3,EMBlackHole4,EMBlackHole5}. In this work however, we use BBHs as a test case as the signals are short and are fast to compute, but will apply our method in the future work for BNS or NSBH events.    
It is well known that each interferometer is more or less an all-sky monitor with very little directional information. Therefore localization isn't possible with a single detector. Having two detectors however does give us some directional information. Using wave arrival information and simple trigonometric calculations, we can get a ring of possible source directions in the sky for a detected GW signal. It is further possible to narrow down this range of possibilities by using relative amplitude and phase information from the detected signals.

 The detectors are most sensitive to waves perpendicular to the plane formed by the two arms of the interferometers. As the incidence angle becomes less than the perpendicular, the sensitivity drops. The curvature of the Earth causes an angle difference of around 27 degrees between the zenith direction of LIGO Hanford and Livingston that creates the aforementioned amplitude and phase inconsistencies. Therefore, it is possible to constrain the sky direction of the source to sufficiently small portions of the sky using only the time delay, phase difference and relative amplitude information. 

The localization can be further improved by using more detectors with large separation baselines between each individual interferometers. For three detectors, LIGO Hanford, LIGO Livingston and Virgo, the signals are localized via a triangulation of the detectors based on observed time delays of the signals. This gives two sky locations that are mirror images with respect the plane of the detectors. The degeneracy between the two sky directions can be broken by using the amplitude and phase information mentioned previously. There are several available algorithms for localising GW signals \cite{localise1,localise2,localise3,localise4,localise5,localise6,localise7,localise8,localise9}. In this work, we use different signal processing techniques to obtain the time delays, amplitude ratios, and phase lags between the three detector signals and use them as input features of our ANN model described later in detail.   
\section{Artificial Neural Networks}\label{Section 4}

As discussed in section \ref{sec:level1}, ANNs \cite{ANN} are machine learning algorithms that are used to perform various tasks, including classification and regression. For this algorithm to work, the network must be `trained' with sufficient labelled examples so that it learns the complex relationships between the input data and the output. The performance of the model is then tested on a set of data which the model has never seen before. A well-trained model will then be able to generalize well on unseen data and give high test accuracy.

The basic processing unit of an ANN is a node or a `neuron', modelled after the neurons in our brain. The ANN is composed of stacked arrays of these neurons that take a vector of inputs ($\vec{x}$) and performs a linear operation on it with some weights ($\vec{w}$) and bias parameters ($b$), which are optimised during the course of training. The output of the transformation operation $f(\vec{x}) = \vec{w}\cdot \vec{x} + b$ from each layer is then passed through a non-linear activation function that restricts the neuron output to a certain range. The commonly used activation functions include rectified linear unit (ReLU), the hyperbolic tangent (tanh), softmax and the sigmoid functions. 
In our work, we have used the ReLU activation function, mathematically expressed as $max(0,x)$.

\begin{figure}[b]
\includegraphics[width=9cm]{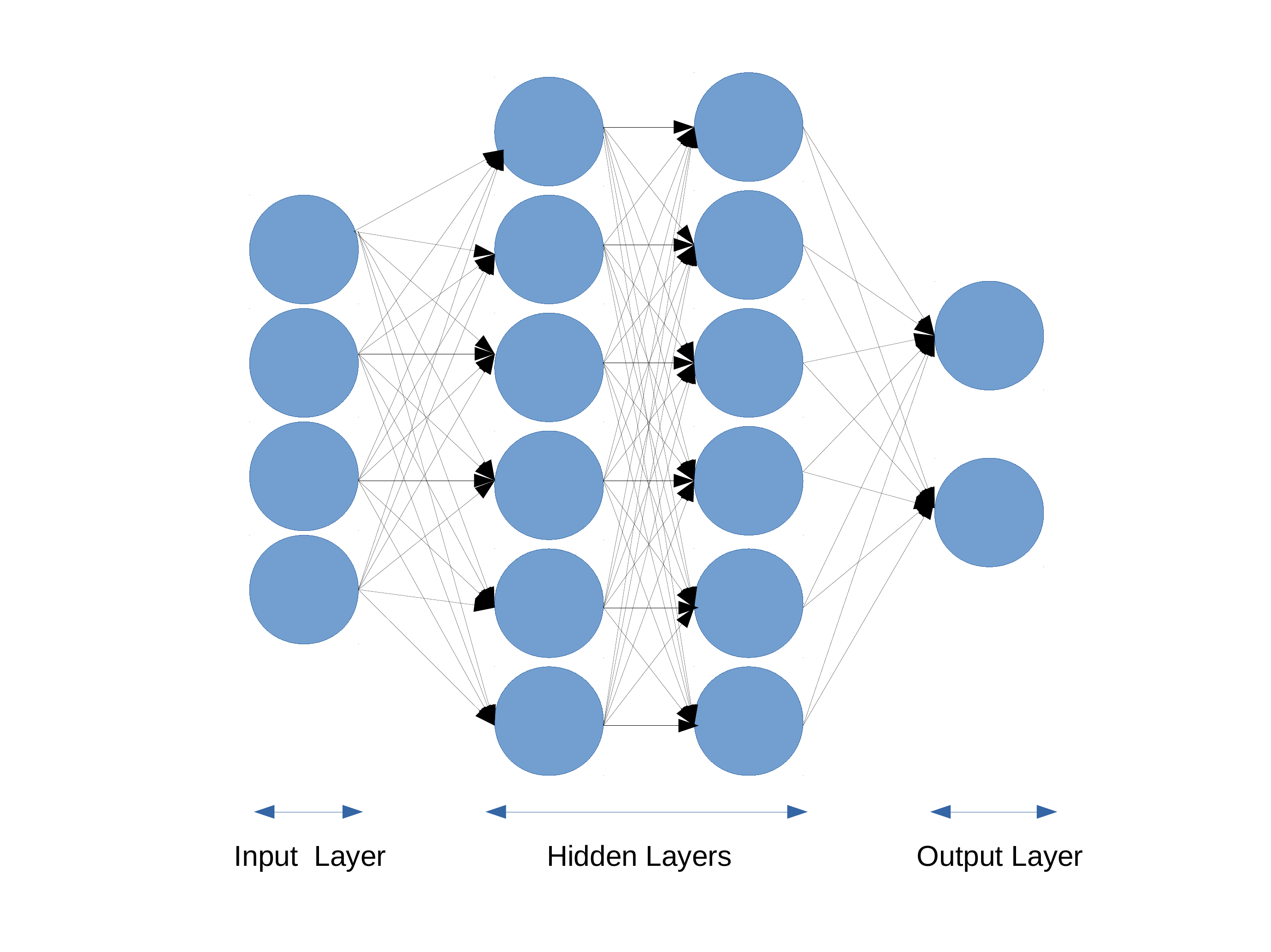}% Here is how to import EPS art
\caption{\label{fig:ANN} An Artificial Neural Network (ANN) with two hidden layers and six nodes in each hidden layer. The circles(neurons) are connected to each other through weights and biases (represented here by arrows).}
\end{figure}

The ANN consists of an input layer of neurons, followed by one or more hidden layers and an output layer as shown in Figure~\ref{fig:ANN}.  Each neuron in these layers is connected to all the neurons in the next layer. In a classification problem, the output at each neuron in the output layer is the probability of an input data to belong to one of the several classes involved in the classification. 

The flow of information from the input layer to the output layer is unidirectional and happens through a process called `forward propagation'. The values obtained at the output nodes are then compared with their true outputs and the losses or mean squared deviations from the true outputs are calculated. To minimise these losses, the errors are back-propagated through the network and the weights and biases are adjusted at each iteration of the training in order to reduce the overall error. This is achieved through algorithms called `gradient descent'. The weights and biases are initialized to small values and are then adjusted through back-propagation over several steps called `epochs', until the errors reach a minimum.

There are several advantages of using ANNs or neural networks in general, over other machine learning algorithms. ANNs have the ability to learn complex non-linear relationships between input data and output and has been found to be able to infer unseen relationships on unseen data. Also, ANNs do not impose any restrictions on the distribution of input variables which is a great advantage over other prediction techniques.

\section{Sample Generation}\label{Section 5}

\begin{figure*}
  \includegraphics[width=17cm]{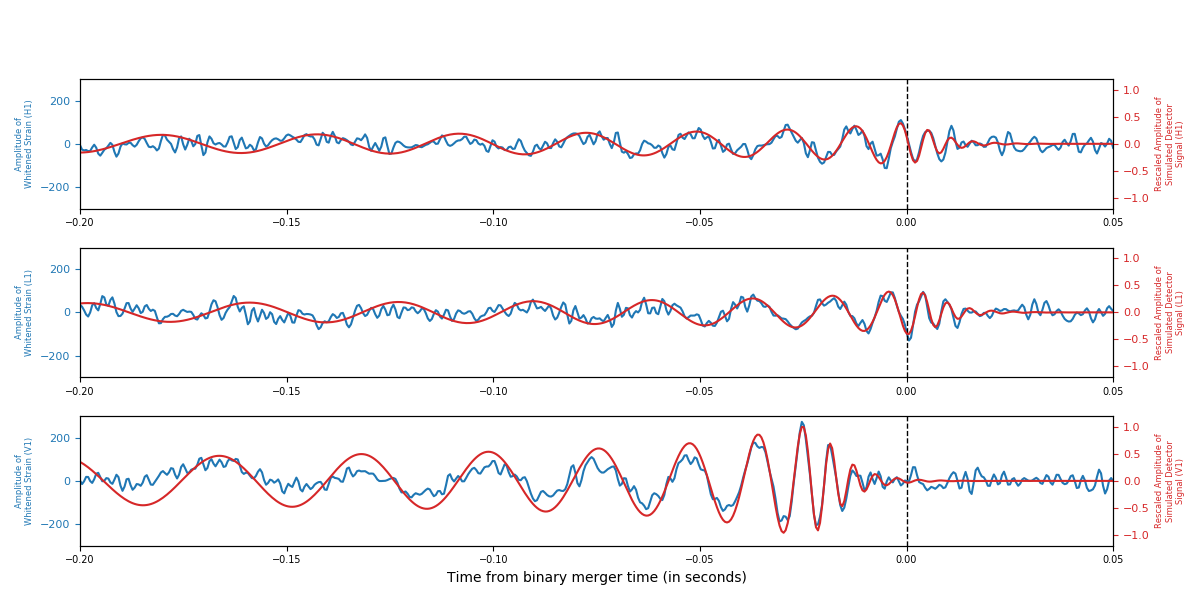}
\caption{\label{fig:1}Gravitational Wave strain}
\end{figure*}

In this section, we give a short description of the sample generation code we used to create the training and testing datasets. We have used a modified version of the code developed by Gebhard et.al. \cite{convwave} for this purpose. While Gebhard et.al. generated GW strains for only LIGO Hanford and Livingston interferometers, we have edited the code to include Virgo as well for better localization of the generated samples.  
Figure~\ref{fig:1} shows an example of a gravitational wave strain sample generated by the code for the three detectors, Hanford, Livingston and Virgo. Here, the pure GW signals, shown in red, were injected into random Gaussian noise and the strain signal, shown in blue was obtained.

To simulate the GW waveforms, we use a joint distribution of a set of relevant parameters as enlisted in Table \ref{tab:table1}. The component black hole masses, spins, polarization angle, cosine of inclination, COA phase angle and injection SNRs are all uniformly sampled from the specified ranges. A brief explanation for each of these parameters and their significance are given in the Appendix in \cite{convwave}.

We concentrate on binary black hole mergers in our classification problem whose waveforms are modelled using the effective-one-body (EOB) approximant \texttt{SEOBNRv4} in the time domain \cite{SEOBNRv4}. The waveforms are produced with a sampling rate of 2048 Hz. The samples are all uniformly distributed throughout the sky with the RA and Decs assigned as follows: RA = 2$\pi u$ and Dec = sin$^{-1}(1-2v)$, where $u$ and $v$ are uniformly sampled between [0,1]. The cosine of the inclination angle is uniformly sampled within the range [-1,1].

These parameters are then fed to routines in LALS\begin{small}UITE\end{small} that performs the actual simulation and returns a tuple with the two polarization modes of the simulated waves ($h_{+}$ and $h_{\times}$), resized to a chosen waveform length. These These `raw' waveforms are then projected onto the antenna patterns of the detectors \cite{antennabook1,antennabook2,antennabook3} based on  their individual RA, Decs and polarization angles using PyCBC \cite{alex_nitz_2019_2581446} functions and are time-corrected to ensure that the signals at the three detectors have the correct time offset on the basis of their relative positions.

This generates the pure signals observed by the detectors without any noise. In this work, we do not take into consideration the effect of the rotation of the Earth on it's axis. We assume that the Earth is stationary and the GW signals are uniformly distributed over the entire sky and approach the detectors from all directions. Since the sensitivity of the network detectors has a functional dependence on the angle of incidence of the GW, this effect is not entirely insignificant. We shall take this into account in our future work.

\begin{table*}%The best place to locate the table environment is directly after its first reference in text
\caption{\label{tab:table1}%
Ranges of values of parameters used to generate the gravitational wave samples.
}
\begin{ruledtabular}
\begin{tabular}{cccccc}
\textrm{Mass 1 and Mass 2\footnote{In solar masses}}&
\textrm{Spin 1 and Spin 2\footnote{z-component of spin}}&
\textrm{Polarization}&
\textrm{COA Phase}&
\textrm{Inclination}&
\textrm{SNR}\\ [0.5ex]
\hline\hline
30-80 & 0-0.998 & [0,2$\pi$] & [0,2$\pi$] & [0,$\pi$] & 50-55\\
\hline
30-80 & 0-0.998 & [0,2$\pi$] & [0,2$\pi$] & [0,$\pi$] & 20-35\\
\hline
30-80 & 0-0.998 & [0,2$\pi$] & [0,2$\pi$] & [0,$\pi$] & 10-110\footnote{With Curriculum Learning}\\
\hline
\end{tabular}
\end{ruledtabular}
\end{table*}

\subsection{Background noise}

\begin{figure}
\includegraphics[width=8cm]{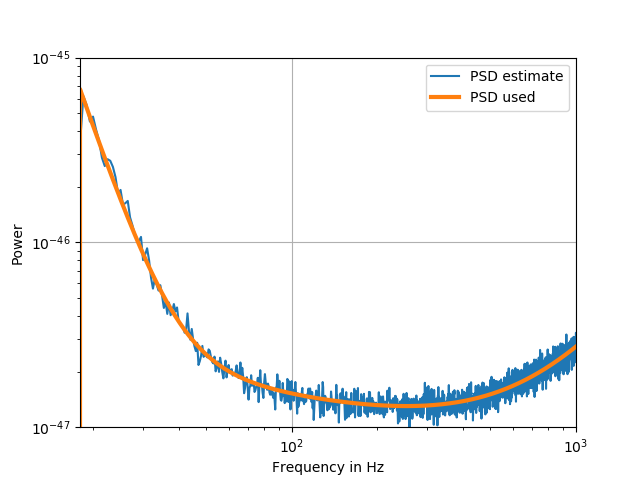}% Here is how to import EPS art
\caption{\label{fig:psd} The Power Spectral Density (PSD) estimated from the background Gaussian noise is shown in blue and the original PSD used to obtain the noise is shown in orange.}
\end{figure}

After the pure signals are generated, they need to be injected into suitable background noise. In this work, we concentrate on simulated Gaussian noise whose frequency distribution is shown in Figure~\ref{fig:psd}. We have used the advanced LIGO PSD, created using the PyCBC module:\texttt{aLIGOZeroDetHighPower}. The simulated waveform is then added to the noise to get the dummy strain. The PSD estimate is obtained from the dummy strain and is then used to calculate the optimal matched filtering signal-to-noise ratio (SNR) for each of the detectors for a particular injection. From this, the network optimal matched filtering SNR (NOMF-SNR) is calculated which is given by :

\begin{equation}
    \sqrt{SNR[H1]^{2} + SNR[L1]^{2} + SNR[V1]^{2}}.
\end{equation}

The NOMF-SNR is used to calculate a rescaling factor that is multiplied to the waveforms to ensure that the resulting waveforms have the chosen injection SNR for the particular sample.

The resulting strain is then whitened and bandpassed, with the lower limit set to 18 Hz and upper limit to 500 Hz. We reject higher frequency components, since for this work, we concentrate on high and intermediate mass black holes. The bandpassed strain is then cut to a length of 0.25 secs, the choice of which is motivated by the estimated inspiral-merger-ringdown length for black hole mergers of our chosen masses and from system memory considerations as well. The distance of all the sources from the Earth is kept fixed at 1000 Mpc. Although the distance scales with the amplitudes and therefore the SNRs of the samples, it is an irrelevant parameter in this context because the waveforms are rescaled to match our chosen injection SNRs, as explained above. More details of the sample generation process can be found in \cite{convwave}.

\subsection{Labelling of generated samples}

\begin{figure*}
\includegraphics[width=14cm]{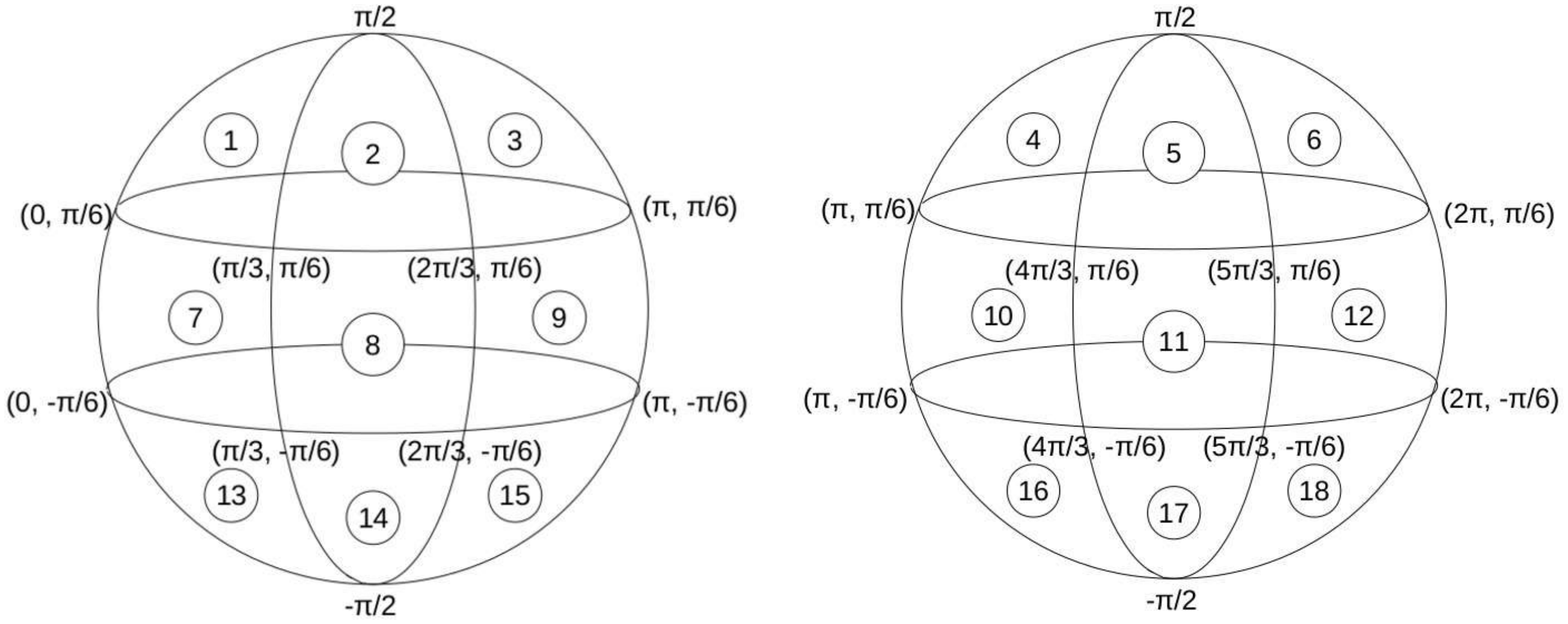}% Here is how to import EPS art
\caption{\label{fig:sectors} Labelling convention for 18 sectors. The RA and Dec ranges for each of the labelled sectors are indicated. We follow similar convention for other sector numbers as well.}
\end{figure*}

Since we are treating the localization of GW signals as a classification problem, we need to first create the classes to which the signals need to be classified into. In order to do this, we model the sky as a sphere and divide it into several sectors based on ranges of RA and sine of Dec. 
We consider cases where the sphere is divided into 18, 50, 128, 1024, 2048 and 4096 sectors. 
Figure~\ref{fig:sectors} shows an example where the sphere has been divided into 18 sectors and labelled according to our defined convention. We have followed a similar labelling convention for all the cases we have considered. 
For each of these cases, we assign labels to the GW samples according to the sectors they belong to. After the model is trained, we test its classification accuracy on a set of GW samples that the model has not been trained with. Each GW sample is assigned only one label. We shall consider the case for multilabelling in a later section. We have used the Pandas library \cite{Pandas} for the purpose of handling data frames and assigning the labels as explained above.  

Here we report the training and test accuracy of classification for the coarse sky resolution (18, 50 and 128 sectors) and use finer resolution (1024, 2048 and 4096 sectors) for testing our model's localization with the parameters of the BBH merger events, GW150914, GW170818 and GW170823. We thereby report and comment on how well our model can localize these events with Gaussian noise and advanced LIGO PSD for three detectors.

\section{Our Method}\label{Section 6}
In this section, we describe the architecture of our ANN and the input features we have used to train the network in the context of localization of GW signals. We explain the method to extract these features from the raw time series data generated by the code described earlier. We consider two different scenarios, pure signals without noise; and signals injected into Gaussian noise. The architectures of our ANN and the input features are tailored for these two cases, which we describe in detail in the following sections. We have written our code in Python 3.6 and have used Keras \cite{Keras} with Tensorflow backend \cite{Tensorflow} to implement our ANN model.

\subsection{Model Architecture}

%\begin{figure}
%\includegraphics[width=10cm]{Model.pdf}% Here is how to import EPS art
%\caption{\label{fig:model} Architecture of our ANN model.}
%\end{figure}
In this section, we describe the architecture of our ANN models. Tables \ref{tab:table2} and \ref{tab:table3} list the components of the two networks for pure signals and signals plus Gaussian noise respectively. The tables consist of three columns. The first column lists the type of layer. `Dense' refers to a fully connected layer, in which each node is connected to all other nodes in the next layer. `Batch Norm' refers to Batch Normalization layer which normalizes the outputs of the previous activation layers by subtracting the batch mean and dividing by the batch standard deviation. It is applied to prevent the network from becoming too sensitive to the initial weights. By including this layer, we reduce the chances of getting very different losses on the test set at each run. The `Dropout' layers ensure that the model does not `overfit' on the training data. Overfitting is a common problem in machine learning, which happens when the model is trained too heavily on the training set, giving a very high training accuracy, but a low test set accuracy. This happens when the model picks up the particular set of details and random fluctuations in the training data too well, but is unable to generalize when presented with new data where these details and fluctuations do not apply. Dropout helps to prevent this over-training by randomly dropping a fraction of the total number of weights in that particular layer. We have used a dropout fraction of 0.2 after each hidden layer in our ANN model for signal plus Gaussian noise. 

Columns 2 in Table I and II lists the output shape of the particular layer. All the layers in our network consist of vectors of dimension equal to the number of nodes in that layer. Column 3 shows the total number of parameters (weights and biases) corresponding to the particular layer. The activation functions we have used in our models are ReLU in all the hidden layers and softmax layer in the output layer. 

\begin{table}%The best place to locate the table environment is directly after its first reference in text
\caption{\label{tab:table2}%
ANN architecture for signal without noise (for 128 sectors). \\
Total parameters: 44,228, 
trainable parameters: 44,228, 
non-trainable parameters: 0 
}
\begin{ruledtabular}
\begin{tabular}{ccc}
\textrm{Layer (type)}&
\textrm{Output Shape}&
\textrm{Parameters} \\ [0.5ex]
\hline\hline
Dense 1 & \multirow{2}{5em}{(None,100)} & 1000\\

Activation 1 &  & 0\\
\hline
Dense 2 & \multirow{2}{5em}{(None,100)}  & 10100\\

Activation 2 &  & 0\\
\hline
Dense 3 & \multirow{2}{5em}{(None,100)}  & 10100\\

Activation 3  &  & 0\\
\hline
Dense 4 & \multirow{2}{5em}{(None,100)}  & 10100\\

Activation 4  &  & 0\\
\hline
Dense 5 & \multirow{2}{5em}{(None,100)}  & 12928\\

Activation 5  &  & 0\\
\hline
\end{tabular}
\end{ruledtabular}
\end{table}

\begin{table}%The best place to locate the table environment is directly after its first reference in text
\caption{\label{tab:table3}%
ANN architecture for signal plus Gaussian noise (for 128 sectors). \\
Total parameters: 195,440,
trainable parameters: 193,184,
non-trainable parameters: 2,256.
}
\begin{ruledtabular}
\begin{tabular}{ccc}
\textrm{Layer (type)}&
\textrm{Output Shape}&
\textrm{Parameters}\\ [0.5ex]
\hline\hline
Dense 1 & \multirow{4}{5em}{(None,200)} & 4400 \\

Batch Normalization 1 &  & 800\\

Activation 1 &  & 0\\

Dropout 1  &   & 0 \\
\hline
Dense 2 & \multirow{4}{5em}{(None,200)} & 40200 \\

Batch Normalization 2 &  & 800\\

Activation 2 &  & 0\\

Dropout 2  &   & 0 \\
\hline
Dense 3 & \multirow{4}{5em}{(None,200)} & 40200 \\

Batch Normalization 3 &  & 800\\

Activation 3 &  & 0\\

Dropout 3  &  & 0 \\
\hline
Dense 4 & \multirow{4}{5em}{(None,200)} & 40200 \\

Batch Normalization 4 &  & 800\\

Activation 4 &  & 0\\

Dropout 4  &   & 0 \\
\hline
Dense 5 & \multirow{4}{5em}{(None,200)} & 40200 \\

Batch Normalization 5 &  & 800\\

Activation 5 &  & 0\\

Dropout 5  &   & 0 \\
\hline
Dense 6 & \multirow{3}{5em}{(None,200)} & 25728 \\

Batch Normalization 6 &  & 512\\

Activation 6 &  & 0\\
\hline
\end{tabular}
\end{ruledtabular}
\end{table}

%Fig.~\ref{fig:model} shows a flowchart of our model architecture.

\subsection{Training Process}

The model is trained after extracting the input vector of features from the time series strain data using several signal processing techniques as described in the next section. Before they are fed into the network, we normalize the features using $(x - \mu)/\sigma $, where $ \mu $ and $ \sigma $ are the mean and standard deviations for all the samples used in the training.

In order to ensure that the model does not overfit, we split our training set into two parts: the training and validation set. The model is evaluated on the validation set at every epoch and the loss on this set is computed to keep track of the generalization efficiency of our model. We also use Keras' EarlyStopping module for this purpose, which monitors the loss on both the training and the validation sets and forcefully ends training when the loss on the validation fails to improve after a certain number of epochs. 

The weights were initialized randomly and during training, were optimized using the stochastic gradient algorithm, Adam \cite{adam}. The loss function used was categorical cross entropy. The calculated loss is back-propagated through the network to updates the weights and biases in order to minimize the loss.

We had generated 12000 samples for pure GW signals without noise. We split it into a training and test set consisting of 10800 samples and 1200 samples respectively. For signals plus Gaussian noise, we have trained the network with 160000 samples and used a validation set of 40000 samples and a test set of 4000 samples. The choice of the number of samples were motivated by both memory considerations and the complexity of the problem. Since it is difficult to make models learn from noisy signals, we used a significantly larger dataset for signal plus Gaussian noise. During the training, we had used a mini-batch size of 250 for pure signals and 2000 for signals plus Gaussian noise. We had experimented with several batch sizes and found these yielded the best results. 

We considered three different SNR ranges in our experiments with signal plus Gaussian noise : [50-55], [20-35] and [10-110]. The third case was studied by applying curriculum learning \cite{curri1} in which the training begins with very high SNRs and then as the training progresses, the SNRs are gradually decreased. This is a form of transfer learning, in which a network uses the tuned weights and biases from earlier training, and applies it to the current training session. This approach helped improve the classification performance at low SNRs while retaining its efficiency at high SNRs. Curriculum learning has already been applied in classification, regression and denoising of GW signals previously  \cite{detection2,detection3,detection5,detection7,denoising1,denoising4,curri2,curri3}.
We describe the results of the classification for all the three cases in later sections. 

\subsection{Input Features}

The most crucial step involved in this work is the choice of robust and effective input features that our ANN has to learn from in order to classify GW signals to their correct sectors. Our choice is directly motivated by existing methods of triangulation of three detectors, as explained earlier. The three most important requirements for accurate localization of GW signals are the arrival time delays of the signals at the three detectors, the phase lags and the amplitudes observed at the three detectors. 
While it is easy to extract this information from signals without noise, the problem becomes much more difficult for noisy signals because information from the actual signal could be lost due to the random fluctuations from terrestrial and other sources. Therefore, on top of the raw time-series strain data, we applied several robust signal processing techniques to extract the features of the signals buried in noise. This method can be generalized to a variety of sources with different black hole masses, spins etc. The techniques are described as follows:

\subsubsection{Cross-correlations of time domain signals}
A common technique in signal processing used to measure the similarity between two signals $u(t)$ and $v(t)$ as a function of the time lag between the them is called cross-correlations, which effectively is a sliding inner product of the two vectors. The cross-correlation between the signals $u(t)$ and $v(t)$ is given by: 
\begin{equation}
w(t) = u(t) \otimes v(t) \overset{\Delta}{=} \int_{-\infty}^{+\infty} u^{*}(\tau)v(\tau + t)d\tau,
\end{equation}
where $u^{*}(\tau) $ denotes the complex conjugate of $u(\tau) $ and $t$ is the time lag between the two signals. Calculating the cross-correlation gives a peak at the time lag where $u(\tau) $ best matches $v(\tau + t) $. 

Since we are dealing with three signals at the three detectors: Hanford, Livingston and Virgo, with different arrival times relative to each other and different amplitudes, we calculate three cross-correlations: between signals at Hanford (H1) and Livingston (L1), Livingston and Virgo (V1) and Hanford and Virgo. The time corresponding to the maximum cross-correlation values give the arrival time delay of one signal relative to the other.

We use the maximum cross-correlation values themselves as the amplitude information necessary to break the degeneracy between two possible sky directions of the source, as explained earlier. Therefore, we use direct cross-correlations of the time series signals to obtain a total of six feature vectors : 

\begin{itemize}
    \item \textit{Arrival time delays} : H1-L1, L1-V1, H1-V1.
    \item \textit{Maximum cross-correlations} :
    H1$\otimes$L1, L1$\otimes$V1, H1$\otimes$V1.
\end{itemize}

\subsubsection{Hilbert transforms}
Since the arrival time delays and maximum cross-correlation values of the time-domain signals are sufficient to estimate the sources of GW signals, we get a very high classification accuracy by running our ANN on pure signals using these features. But for noisy signals, the accuracy drops significantly because of the distortion of the signal due to the randomness of the noise. Therefore, besides the cross-correlations, we make use of the analytic representations obtained through a Hilbert transformation \cite{hilbert1,hilbert2}. 
The Hilbert transformation converts a time domain signal into a vector of complex numbers and allows us to easily extract quantities like the instantaneous amplitudes, instantaneous phases and phase lags between two signals, from noisy signals in the time-domain.

The analytic signal of $s(t)$ with its Fourier transform $S(f) $ is given by :
\begin{equation}
s_{a}(t) \overset{\Delta}{=} \mathcal{F}^{-1}[S(f) + sgn(f).S(f)], 
\end{equation}
where sgn($f$) is the signum function and $ \mathcal{F}^{-1} $ is the inverse Fourier transform. 

This can be expressed as:
\begin{equation}
\mathcal{F}^{-1}\{ S(f) \} + \mathcal{F}^{-1}\{sgn(f) \}*\mathcal{F}^{-1}\{S(f)\},
\end{equation}

where * denotes convolution operation. From here we get, 
\begin{equation}
 = s(t) + \underbrace{j[\dfrac{1}{\pi t}* s(t)]}_{\mathcal{H}[s(t)]},
\end{equation}
where $\mathcal{H}[s(t)]$ is the Hilbert transform of $ s(t) $.

The analytic forms of the detector signals (H1$_{a}$, L1$_{a}$, V1$_{a}$) are therefore arrays of 512 complex numbers (all the GW samples generated are 0.25 s long with a sampling frequency of 2048 Hz) whose real part gives the \textit{instantaneous amplitude} and complex part, the \textit{instantaneous phase}. 

We used cross-correlations of the analytic signals to obtain the arrival time lags and amplitude information from the three pairs of detectors, the same way as we had done for the time domain signals. 

We calculate the \textit{average instantaneous phase around the merger time} of the three signals from the complex part of the Hilbert transform and find the difference of these average phases for each pair of detectors. We first find the instantaneous amplitudes around the merger region. By searching for ten highest peaks from each GW sample for the three detectors and record their instantaneous phases. Then we calculate the average of the ten instantaneous phases. This gives us the average instantaneous phase around the merger of each signal. We then calculate the differences of these instantaneous phases for each pair of detectors to get the \textit{instantaneous phase lags}. 

Having already found the instantaneous amplitudes around merger for each signal, we calculate their average and take the ratios for each pair of detectors. This serves as an additional feature for the ANN. The importance of using features around merger time is that since the signal is loudest in this region, the effect of noise is minimal, especially for high SNRs. Hence the ANN is able to learn the characteristics of the pure signal much better than while using the entire signal.

We therefore obtain 12 more input features from this method of using analytic signals via Hilbert transforms:
\begin{itemize}
    \item \textit{Arrival time delays}: H1$_{a}$-L1$_{a}$, L1$_{a}$-V1$_{a}$, H1$_{a}$-V1$_{a}$
    \item \textit{Maximum cross-correlations}:  H1$_{a}\otimes$L1$_{a}$, L1$_{a}\otimes$V1$_{a}$, H1$_{a}\otimes$V1$_{a}$.
    \item \textit{Average phase lags}: H1$_{a}$-L1$_{a}$, L1$_{a}$-V1$_{a}$, H1$_{a}$-V1$_{a}$
    \item \textit{Ratios of average amplitudes around merger}: H1$_{a}$/L1$_{a}$, L1$_{a}$/V1$_{a}$, H1$_{a}$/V1$_{a}$.
\end{itemize}
Note, to compute the analytic signals using the Hilbert transforms, we used Scipy's \cite{Pandas} \texttt{signal.hilbert} method.

\subsubsection{Complex correlation coefficients}

We discussed earlier that for the purpose of localising the GW signals, the three interferometers are \textit{triangulated}, which means while operating together, they form a two dimensional plane in space. GW signals may arrive at the interferometers at any angle with respect to that plane. However, since the interferometers themselves are located some distance from each other, the curvature of the Earth will introduce slight differences in the angles between the signals received at the three detectors. This quantity can therefore also be thought of as a measure of the phase difference of the waves. 
We calculate the angles between the signal vectors and use them as the final input feature for our ANN. To calculate the angles, we compute the complex correlation coefficient of each pair of signals in the following way: 
For complex-valued sinusoids, for example, of the form $x(t) = Ae^{j(\omega t+ \psi)} $ and  $y(t) = Be^{j(\omega t+ \phi)} $, the angle $\theta $ between them is given by: 
\begin{equation}
    \theta = arc cos \bigglb( \dfrac{\sum_{n=0}^{N-1} x[n](y[n])^{*}}{\sqrt{\sum_{n=0}^{N-1}|x[n]|^{2}}\sqrt{\sum_{n=0}^{N-1}|y[n]|^{2}}} \biggrb).
\end{equation}
where $x$ and $y$ have $N$ samples each. 
We perform this operation on H1, L1 and V1 signals using Scipy's \cite{Pandas} libraries and get the final set of input features for signals + Gaussian noise:

\begin{itemize}
    \item \textit{Complex correlation coefficients} =  H1$*$L1, L1$*$V1, H1$*$V1.
\end{itemize}

\section{Experiments and Results}\label{Section 7}

\subsection{Signal without noise}

Our first experiment was with pure signals. 
We used this as a sanity check for the validity of our approach, since the ANN is certainly not expected to classify noisy signals correctly if it fails for signals without noise. Since the GW signals are not contaminated by noise, we only use input features from the actual signals rather than their analytic representations. These features are the following:
 
\begin{itemize}
  \item Arrival time difference at each pair of detectors.
  \item Relative phase differences between each pair of signals.
  \item Ratios of amplitude at merger time. (taken as the time corresponding to highest amplitude of the signal).  
\end{itemize}
 
We generated a sample of 12000 GW signals, distributed uniformly over the sky and split it into the training and test sets having 10800 and 1200 samples respectively. We used fixed masses, spins, COA phase angles, inclinations and polarization angles for these experiments. The architecture of the ANN we used is shown in Table \ref{tab:table2}. We used a mini-batch size of 250 and trained the ANN for 300 epochs. We repeated this experiment for three different sector numbers: 18, 50 and 128. The training and test accuracy for the three cases are shown in Table \ref{tab:table4}.
 
 We see that both the training and test accuracy are very high for all the three cases, implying our model is able to learn the relationships of the input vector of features with the individual sectors. It is also able to generalize well on unseen data from the test set, resulting in the high test accuracy. We observe that the test accuracy drops as the number of sectors is increased. This is because as the sectors become finer, it gets more and more difficult for the model to pick the correct sector from the ones in its immediate vicinity since the arrival times lags, phase lags and relative amplitudes of the GW signals from the correct sector and those around it become more and more similar in value for closely separated regions of the sky. Besides, the angular resolution is also limited by the SNRs and timing resolution. It must also be noted that this test accuracy corresponds to the number of GW samples that are classified `exactly' correctly. In the next section, for signals plus Gaussian noise, we consider a second measure of test accuracy where samples that are classified to the immediate neighbouring sectors are counted as correctly classified. We find that the test accuracy we obtain by implementing this scheme is above 90\% for all our experiments. 
 
  \begin{table}
 \caption{\label{tab:table4}%
Results for pure signals and coarse angular resolution. Training with 10800 samples, tested with 1200 samples. Batch size = 250, number of epochs = 300.
}
 \begin{ruledtabular}
\begin{tabular}{ccc}
\textrm{Number of sectors}&
\textrm{Training accuracy}&
\textrm{Test set accuracy} \\ [0.5ex]
\hline\hline
18 & 98\% & 95\%\\
\hline
50 & 97\% & 91\%\\
\hline
128 & 95\% & 85\%\\
\end{tabular}
\end{ruledtabular}
\end{table}
 
 \subsection{Signal with Gaussian noise}
 
 Table \ref{tab:table3} shows the architecture of our ANN for signal plus Gaussian noise. The network here is deeper with greater number of nodes in each hidden layer to allow the ANN to learn characteristics of the actual signal from the noisy background. To make the training better, we generated 200000 GW samples of which we used 160000 samples for training and 20000 samples for validation. The ANN was tested on a final, unseen, 4000 samples. We generated our samples with variable component masses, spins etc, randomly sampled within the range as shown in Table \ref{tab:table1}. We considered two different SNR ranges : [50-55] and [20-35]. For each of these SNR ranges, we have evaluated our model on 18, 50 and 128 sectors each. 
 
 In addition, in order to improve the performance of the algorithm at low SNRs, we used the curriculum learning method, explained earlier, where we begin training with high SNR GW samples and then gradually decrease the SNR to 10. We evaluated the performance of our ANN, trained with the curriculum learning approach, using a test set of 4000 samples of SNR in the range of [20-30] and found a 3\% increase in accuracy for 128 sectors. 
 
 We have used the full set of 21 input features for our experiments with noisy signals, which are the following: 
 
 \begin{itemize}
     \item Arrival time delays of signals.
     \item Maximum cross-correlation values of signals.
     \item Arrival time delays of analytic (signal plus Hilbert transform of the signal) signal.
     \item Maximum cross-correlation values of analytic signals.
     \item Ratios of average instantaneous amplitudes around merger.
     \item Average phase lags around merger.
     \item Complex correlation coefficients between the three signals.
 \end{itemize}
 As explained earlier, the test accuracy in all the cases improves once we consider the GWs that are classified into the immediate neighbouring sectors are counted as correctly classified. This proves that our model is indeed reliable.
 
 \begin{table*}[hbt!]
 \caption{\label{tab:table5}%
Results for signals plus Gaussian noise and coarse angular resolution. Trained with 160000 samples, validated with 40000 samples, tested with 4000 samples. Batch size = 2000. 
}
 \begin{ruledtabular}
\begin{tabular}{ccccc}
\textrm{SNR}&
\textrm{Number of sectors}&
\textrm{Training accuracy}&
\textrm{Test accuracy}&
\textrm{Revised test accuracy\footnote{correct within one sector}}\\ [0.5ex]
\hline\hline
[50-55]& 18 & 89\% & 91\% & 98.5\%\\

[50-55]& 50 & 80\% & 84\% & 98.25\%\\

[50-55]& 128 & 70\% & 77\% & 97.8\%\\
\hline
[20-35]& 18 & 80\% & 85\% & 97.27\%\\

[20-35]& 50 & 69\% & 73\% & 96\%\\

[20-35]& 128 & 55\% & 62\% & 92\%\\
\hline
[10-110]\footnote{With Curriculum Learning}& 128 & 60\% & 65\% & 94.5\%\footnote{Tested on samples with SNR of [20-35]}\\
\end{tabular}
\end{ruledtabular}
\end{table*}

\begin{figure*}[h!]
\centering
\subfigure[]{
\includegraphics[scale=0.8]{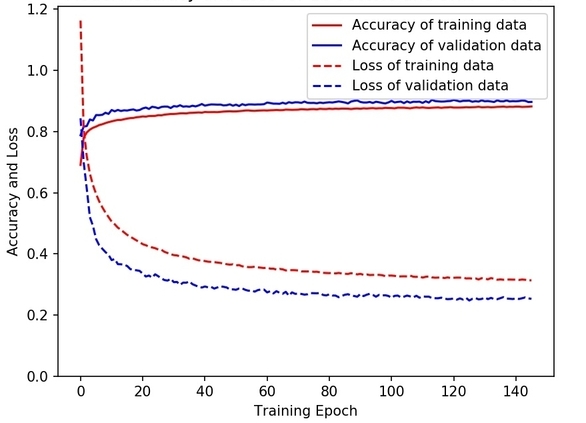}}
\subfigure[]{
\includegraphics[scale=0.8]{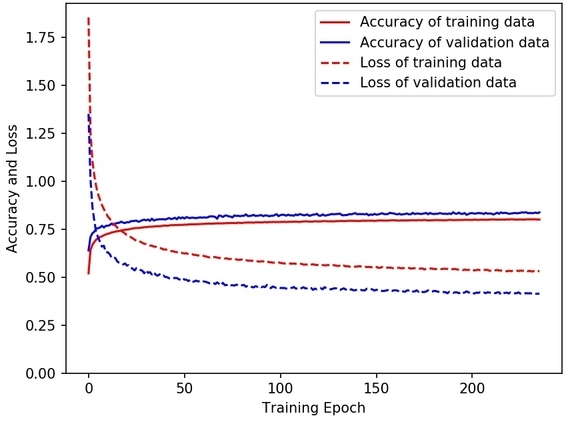}}
\subfigure[]{
\includegraphics[scale=0.8]{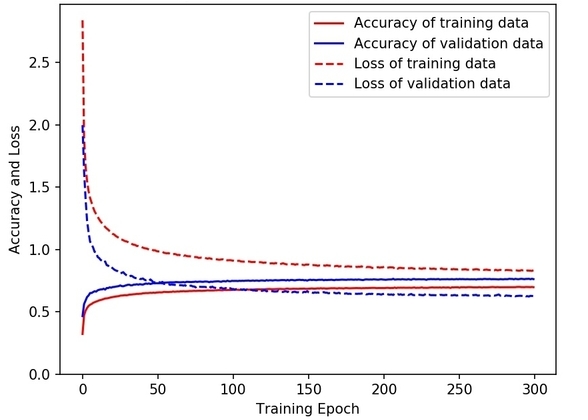}}
\caption{\label{fig:trainingcurve1} Accuracy and loss curves vs. training epoch for (a) 18, (b) 50 and (c) 128 sectors and SNR range of [50-55]  }
\end{figure*}

\begin{figure*}[h!]
\centering
\subfigure[]{
\includegraphics[scale=0.8]{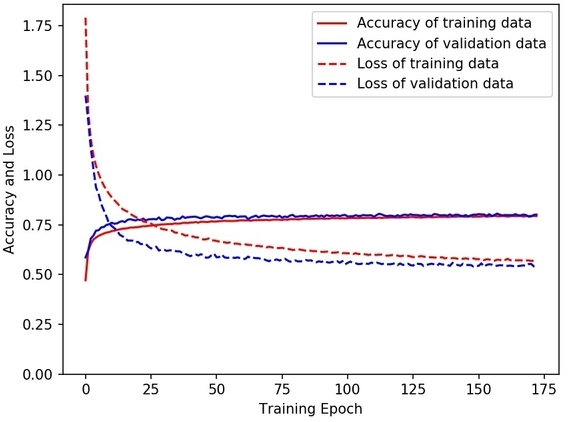}}
\subfigure[]{
\includegraphics[scale=0.8]{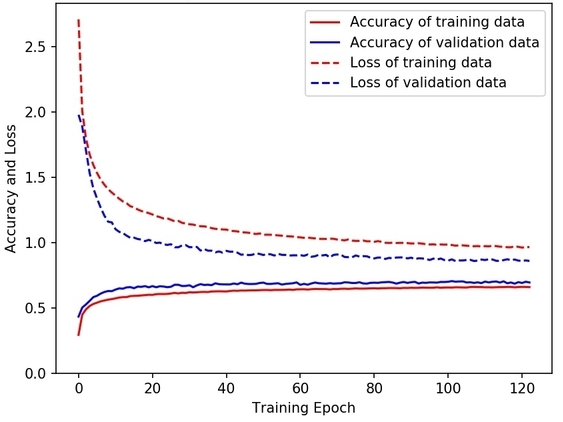}}
\subfigure[]{
\includegraphics[scale=0.8]{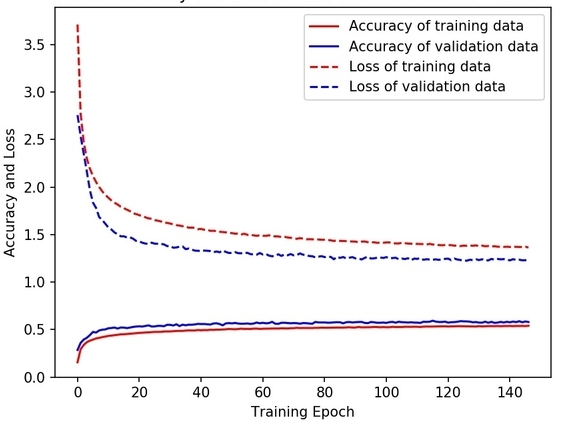}}
\caption{\label{fig:trainingcurve2} Accuracy and loss curves vs. training epoch for (a) 18, (b) 50 and (c) 128 sectors and SNR range of [20-35]  }
\end{figure*}

Figures \ref{fig:trainingcurve1} and \ref{fig:trainingcurve2} show the accuracy and loss curves in the training of the network for 18, 50 and 128 sectors for SNR ranges of 50-55 and 20-35 respectively. From the training and validation loss curves, we can conclude that our model is a good fit for the data for all the cases that we have considered. 
Both the training and validation loss curves in each of the Figures \ref{fig:trainingcurve1} and \ref{fig:trainingcurve2} decreases to a point of stability after the first few epochs after which they remain almost constant. 
The fact that our model does not overfit on the training data is confirmed by the validation loss curve remaining constant until the last epoch and it does not show an inflection point. 
If the model had started to overfit, the losses in the validation set would have climbed after reaching a minimum. Note the usual or preferable scenario is to always have a training loss curve lower than the validation loss curve.
The fact that our validation loss is lower than the training loss curve in our plots suggests that we have an unrepresentative validation set, meaning that the validation dataset does not provide sufficient information to evaluate the ability of the model to generalize. 
This is expected because of the large disparity in the volume of the training and validation sets. 
While we have 160000 training examples, there are only 40000 validation samples. 
However since we have established that the model does not overfit, we can safely ignore this behaviour.
 
\section{Evaluation of performance}\label{Section 8}

We observe that the training and the test set accuracy of our ANN is high for both pure and noisy signals, indicating that our model is robust and is able to generalize well on examples not used in training. However, as the number of sectors are increased, the accuracy for both the training and test sets decrease. 
This is because as the area of the sectors gets smaller, it becomes more and more difficult for the model to differentiate between GW signals coming from adjacent sectors. 
Signals from adjacent sectors, would have nearly the same arrival time delays, phase lags and relative amplitudes when the individual sectors are small. Therefore the model is confused between the `correct' sector and the ones lying adjacent to it, thereby decreasing the accuracy. 
In order to confirm this, we picked those incorrectly classified GW signals from our test set which were located at the sectors adjacent to the correct one and included them with our correctly classified samples. The revised test accuracy we obtained in doing so was greater than 90\% for all our test cases, thereby proving our hypothesis and establishing the efficiency of our ANN algorithm. The last column of Table \ref{tab:table5} shows this revised test accuracy.
 
Fig~\ref{fig:2} shows the performance of the ANN network on a test set comprising 4000 simulated gravitational wave samples generated with SNRs uniformly sampled between 20 and 35, injected into Gaussian noise. These samples are distributed uniformly over the entire sky and has been classified into 50 sectors based on their RA and Dec values, according the convention described earlier. The colored patches in the sphere represent two of those 50 sectors. The dots are the actual sky directions of the GWs plotted at their exact RA and Decs. The blue dots represent GW samples that the ANN classifies into the coloured sectors. Therefore, the blue dots lying inside the coloured patches represent samples correctly classified by the network and those outside it are incorrectly classified. We observe that the actual sky directions of almost all of these incorrectly classified signals are very close to the sectors the ANN classifies them into, thereby indicating that our model is indeed very accurate and is therefore a confirmation of our high `Revised Test Accuracy Scores' in Table V.
 
The orange dots represent samples which belong to the coloured sectors, but were classified to the immediate neighbouring sectors by our ANN. The incorrect sectors assigned to these GW samples by our ANN lie just adjacent to the sectors they actually belong to, reaffirming our ANN's high classification accuracy. 
  
Instead of counting the number of misclassified GW signals lying in the sectors adjacent to the correct one, we can directly implement Keras's `multilabelling' scheme in our code for the same purpose. The idea behind multilabelling is to assign more than one labels or in this case, sectors, to each of our GW samples. Therefore each GW sample can be labelled with the `exactly correct' sector ID as well as with the IDs of all the sectors adjacent to it. This will not only result in high test accuracy, as we have demonstrated, but high training accuracy as well.

\section{Test Result for GW150914, GW170818 and GW170823-like Sources}\label{Section 9}
In Section VIII we have established that our model is able to correctly classify GW samples not used in training with more than 90\% accuracy for SNRs $\geq$ 20 and number of sectors $\leq$ 128. In this section, we investigate how well our model is able to localize the O1 event GW150914 and the O2 events GW170818 and GW170823.
Note our training samples are generated with advanced LIGO PSD for three detectors since the purpose of this investigation is a sanity check to test the feasibility of our approach and to confirm whether in an ideal situation, our method gives good results.

In order to test this, we first simulate the GW signals with the published BBH parameters as listed in Table \ref{tab:EventParameters} for GW150914 and Tables \ref{tab:EventParameters1} and \ref{tab:EventParameters2} for GW170818 and GW170823 events. Next, we consider a large number of sectors: 1024, 2048, 4096 and train our model with samples of SNR in the range of [20-25] for the GW150914 test sample and with SNR in the range of [10-15] for GW170818 and GW170823 test samples. We then test our model's classification on them and use the probability assigned to each sector by our ANN for these three test samples as ranking statistics from which we obtain the cumulative frequency distribution of the probabilities and get the 90\% and 50\% contours in the same way as in \cite{LeoSinger}.

\begin{figure*}[h!]
\centering
\includegraphics[scale=0.7]{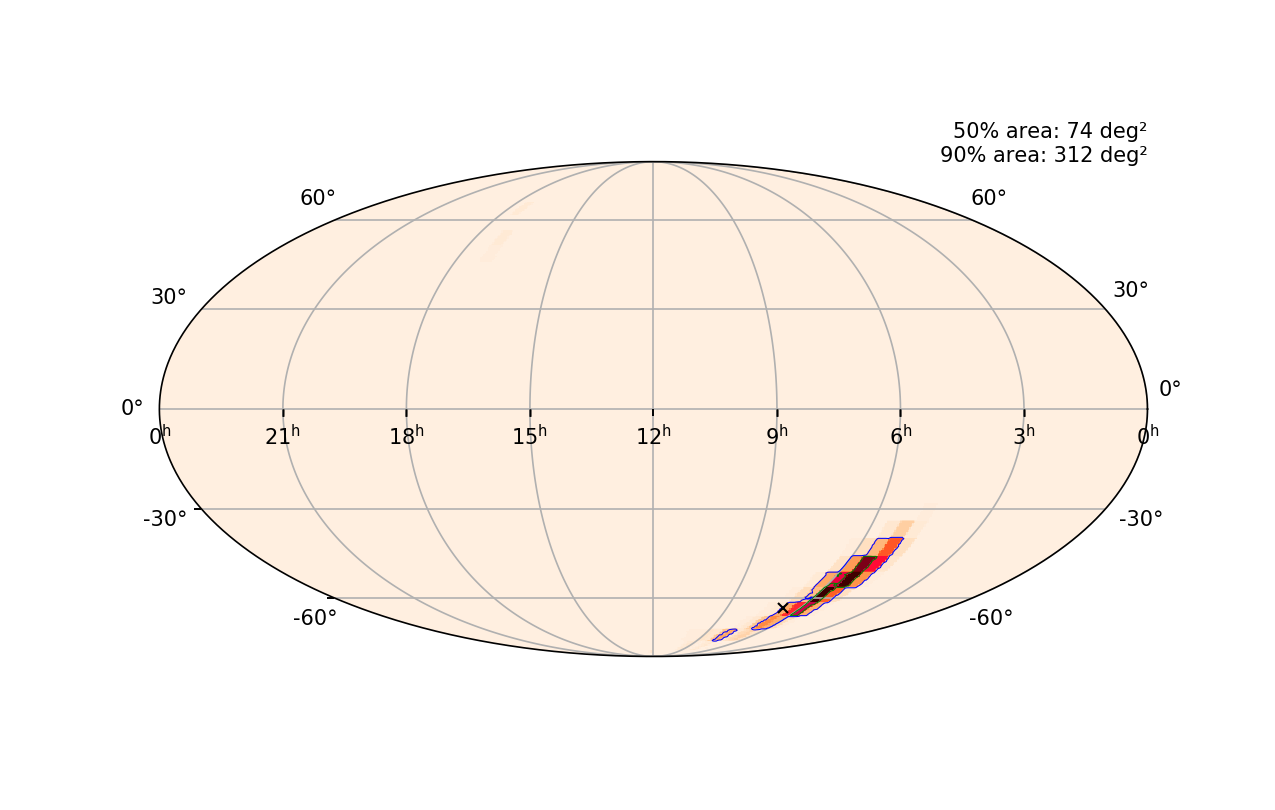}% Here is how to import EPS art
\caption{\label{fig:GW150914} Probability heatmap of localization by our ANN for GW150914 test sample. The blue line shows the 90\% contour and the green line shows the 50\% contour.
The exact sky direction of the GW signal, as chosen by us is marked with a black cross.}
\end{figure*}

\begin{figure*}[h]
\centering
\subfigure[]{
\includegraphics[scale=0.7]{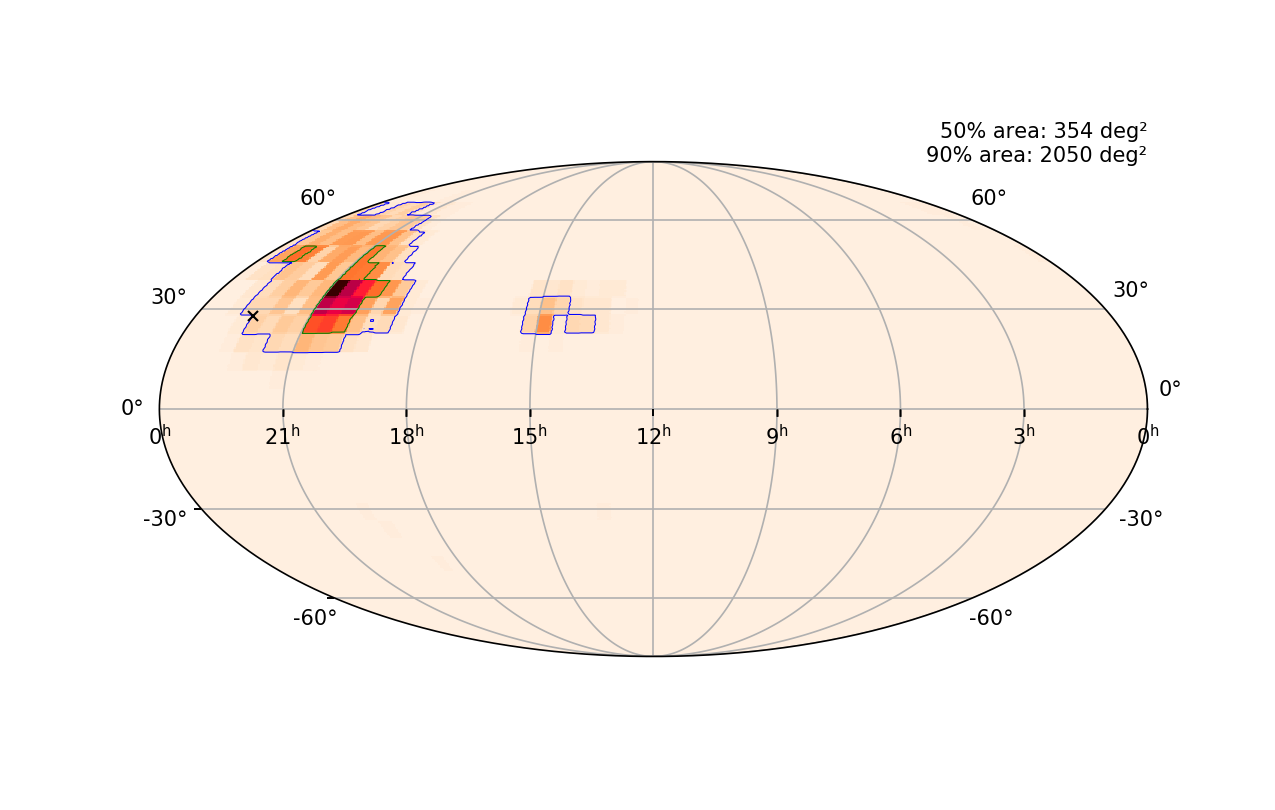}
}
\subfigure[]{
\includegraphics[scale=0.7]{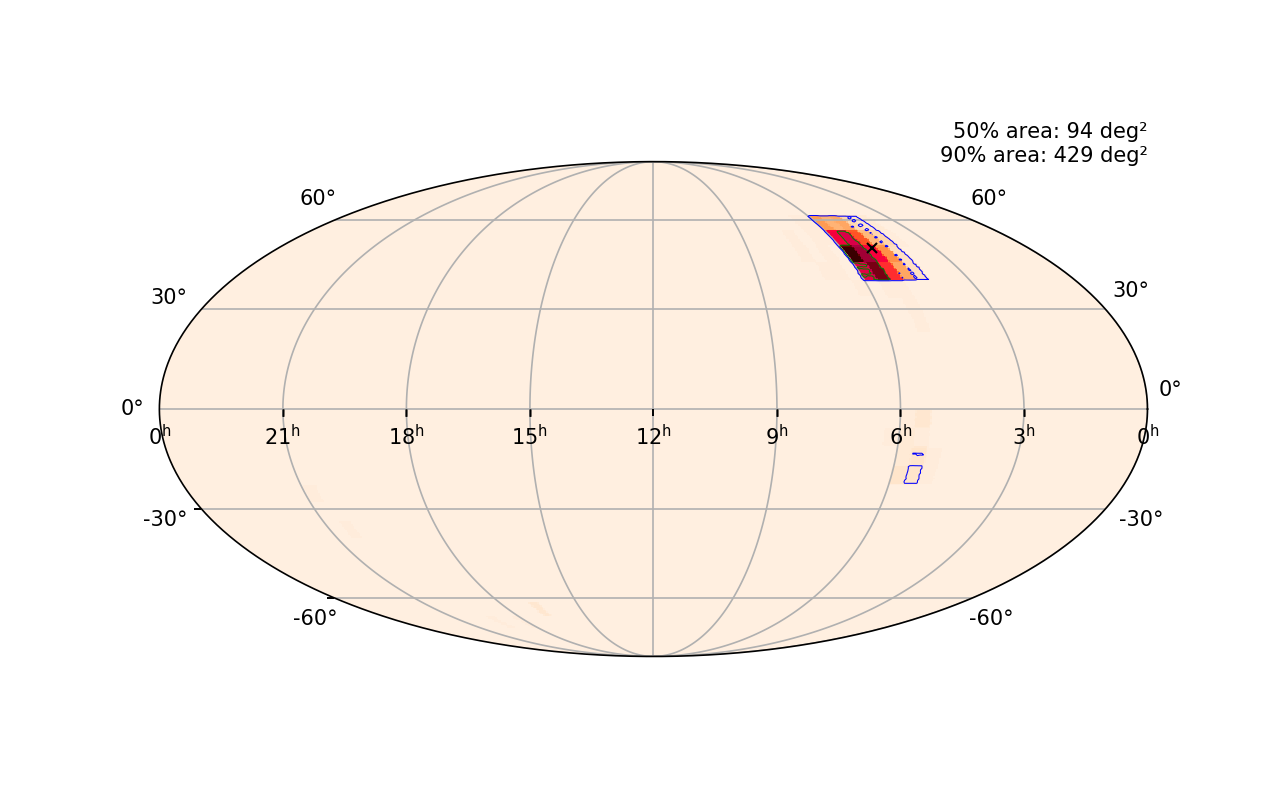}
}
\caption{\label{fig:GW170818_GW170823} (a)Probability heatmap with of localization with contours by our ANN for GW170818 test sample. (b) Probability heatmap with of localization with contours by our ANN for GW170823 test sample. The blue line shows the 90\% contour and the green line shows the 50\% contour.
The exact sky direction of the GW signal, as chosen by us is marked with a black cross.}
\end{figure*}

Figures \ref{fig:GW150914} and \ref{fig:GW170818_GW170823} (a) and (b) show probability heatmaps and confidence intervals respectively, for our test samples with 2048 sectors. They have been plotted using the Python package `healpy', that is based on the Hierarchical Equal Area isoLatitude Pixelization (HEALPix) library. For plots generated using healpy, the sky is divided into `N' pixels. The size and number of pixels depend on our preferred resolution. Higher resolution skymaps have finer and greater number of pixels. Each of these pixels have a specific RA and Dec and a probability associated with it for the event to be located in that particular pixel or coordinate in the sky. Since the sectors in our definition have a finite area and is not simply a point or pixel in the skymap, we assign all the pixels within our sectors the same value of probability equal to the probability that was assigned to the sector by our model. Heatmaps are then obtained from these probabilities, with darker colours denoting higher probability. The 50\% and 90\% contour intervals are then plotted from the cumulative distribution of these probability values. The plots were made using ligo.skymaps software developed by Leo P. Singer \cite{LeoSinger}.

\begin{table}%The best place to locate the table environment is directly after its first reference in text
\caption{\label{tab:EventParameters}%
Intrinsic parameters and SNR used for GW150914 as published in \cite{bp2}}
\begin{ruledtabular}
\begin{tabular}{ccc}
\textrm{Parameters}&
\textrm{Value}&
\textrm{Upper/Lower error}\\
\hline\hline
         
Primary BH mass & 36.2 M$\odot$ & +5.2 -3.8\\
\hline
Secondary BH mass & 29.1 M$\odot$ & +3.7 -4.4 \\
\hline
Primary BH spin & 0.68 & +0.05 -0.06\\
\hline
Secondary BH spin & 0.80 & +0.05 -0.06\\
\hline
SNR & 24 & -\\

\end{tabular}
\end{ruledtabular}
\end{table}

\begin{table}%The best place to locate the table environment is directly after its first reference in text
\caption{\label{tab:EventParameters1}%
Intrinsic parameters and SNR used for GW170818 as published in \cite{GW170818}}
\begin{ruledtabular}
\begin{tabular}{ccc}
\textrm{Parameters}&
\textrm{Value}&
\textrm{Upper/Lower error}\\
\hline\hline
         
Primary BH mass & 35.5 M$\odot$ & +7.5 -4.7\\
\hline
Secondary BH mass & 26.8 M$\odot$ & +4.3 -5.2 \\
\hline
Final spin & 0.67 & +0.07 -0.08\\
\hline
SNR & 11.3 & -\\

\end{tabular}
\end{ruledtabular}
\end{table}

\begin{table}%The best place to locate the table environment is directly after its first reference in text
\caption{\label{tab:EventParameters2}%
Intrinsic parameters and SNR used for GW170823 as published in \cite{GW170818}}
\begin{ruledtabular}
\begin{tabular}{ccc}
\textrm{Parameters}&
\textrm{Value}&
\textrm{Upper/Lower error}\\
\hline\hline
         
Primary BH mass & 39.6 M$\odot$ & +10.0 -6.6\\
\hline
Secondary BH mass & 29.4 M$\odot$ & +6.3 -7.1 \\
\hline
Final spin & 0.71 & +0.08 -0.10\\
\hline
SNR & 11.5 & -\\

\end{tabular}
\end{ruledtabular}
\end{table}

\begin{table}%The best place to locate the table environment is directly after its first reference in text
\caption{\label{tab:ComparisonTable1}%
Areas of 50\% and 90\% confidence contours for our GW150914, GW170818 and GW170823 test samples with 2048 sectors.}
\begin{ruledtabular}
\begin{tabular}{ccc}
\textrm{Event}&
\textrm{90\% contour (in deg$^{2}$)}&
\textrm{50\% contour (in deg$^{2}$)}\\
\hline\hline
         
GW150914 & 312 & 74 \\
\hline
GW170818 & 2050  & 354  \\
\hline
GW170823 & 429  & 94  \\
\end{tabular}
\end{ruledtabular}
\end{table}

We can see from Figures \ref{fig:GW150914} and \ref{fig:GW170818_GW170823} that our model has been able to localize the signal quite well, implying our method is feasible for Gaussian noise and advanced LIGO PSD. With real LIGO noise and two detectors, Hanford and Livingston, Bayestar and PE techniques could localise the 90\% contour of GW150914 to an area of 610 deg$^2 $ and 230 deg$^2 $ respectively \cite{bp1}. With the O3 PSD and Gaussian noise, but with the same network SNR of 24, we can localize the same signal to 312 deg$^2 $. For GW170818, the 90\% area obtained by PE was 39 deg$^2 $ in comparison with our 2050 deg$^2 $. \cite{GW170818}. The reason for this large difference, we believe, is because of insufficient training data in the sectors around the source location. It must also be kept in mind that our localization is limited by our choice of sampling rate of 2048 Hz for our generated signals. With higher sampling rate, we can obtain the time delay with greater precision and therefore localize our source better. We shall take into consideration the effect of higher sampling rate in our future work. Another probable reason may be the unreliability of some of our extracted features at low SNRs. For example, we assume that information like the amplitude ratios around merger regions that we use as one of our feature vectors becomes ambiguous at SNRs < 12. However, we also find that at the same SNR, the 90\% area of our GW170823 test sample is 429 deg$^2 $ in comparison with Bayestar's area of 2145 deg$^2 $ and PE's area of 1651 deg$^2 $ \cite{GW170818}.

One of the major advantages of our model, and also of all ML algorithms in general is that once the model is trained, it can perform classifications or any other operation on the test set at very high speeds, which is of particular importance for EM follow-up, as discussed earlier. 
We find that our trained model is able to perform the classification on a single test sample in 0.0003 secs while running on 14 Intel Xeon CPU cores. 

The total time of execution, taking into consideration the time to extract the input features from the test sample's raw strain data and classifying it into one of 8192 sectors is 0.0189 secs on 14 CPU cores. 
Therefore our model is possibly orders of magnitude faster than any other existing localization techniques, since Bayestar takes 10 secs - 1 min and other offline PE algorithms take hours to perform the localization. 

\begin{figure*}
\includegraphics[width=12cm]{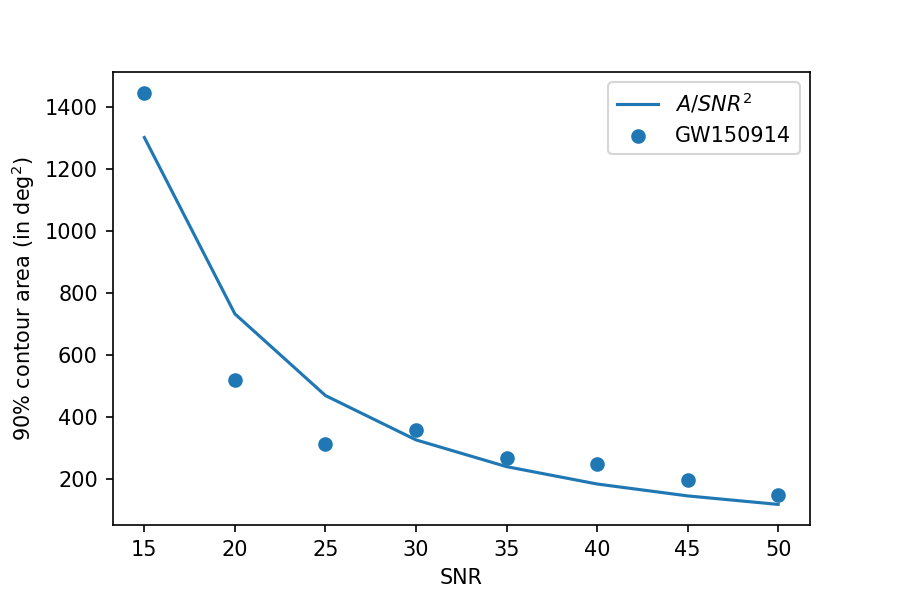}% Here is how to import EPS art
\caption{\label{fig:Convergence} Localization accuracy of ANN vs.SNR for our GW150914 test sample.}
\end{figure*}

We have also performed a convergence test of our localization method on the GW150914, GW170818 and GW170823 test samples for SNR in the range of 15 to 50. Figure \ref{fig:Convergence} shows our convergence plot of 90\% contour area vs. SNR for the GW150914 test sample. This area is expected to decrease as the SNR is increased by following an inverse square relation, as expected from diffraction limit \cite{Wen&Chen}. The curve has been fitted with the function $A/x^{2}$. We find that the 90\% areas of our test sample does follow this curve as the SNR is increased. We have performed this convergence test for SNR in the range of 10 to 50 for our GW170818 and GW170823 test samples as well and we find that the 90\% areas follow a similar trend.  

For all three of our test samples, the 90\% areas decreased as we increased the number of sectors from 1024 to 2048 but slightly increased when we increased the sector number to 4096. We suspect that this is due to the fact that with finer sectors, our training samples are distributed over a larger number of sectors and hence there aren't enough training examples in the sectors around the event to enable the model to accurately learn the features from that part of the sky. As discussed previously, the angular resolution of the source is also limited by our choice of timing resolution of 1/2048 s. Besides, in a classification problem, as the number of classes are increased, the number of error values that are backpropagated through the network during the training also increases and the weights and biases are updated every time. This might lead to instability if the number of classes (sectors in this case) are too large. Since the network may not be tuned to handle so many optimization operations in each epoch of the training, it may affect its performance.

\section{Discussion and Conclusion}\label{Section 10}
 
We have attempted to localize GW signals injected into Gaussian noise using deep learning. We have built an ANN and have approached it like a classification problem, where the sky, modelled as a sphere has been divided into several sectors and we train our ANN model so that it classifies simulated GW signals into their correct sector based on it's RA and Dec's. We report high ($>$ 90\%) classification accuracy for all the cases with coarse angular resolution down to a few hundredths of the sky. To achieve finer angular resolution, we calculate confidence levels using ML probability as the detection statistics. We tested our model's localization accuracy by using test samples with injection parameters of GW150914, GW170818 and GW170823 and conclude that our model gives feasible results for Gaussian noise and advanced LIGO PSD. We demonstrate that the 90\% contour area for our test samples decrease in an inverse proportion to the square of SNRs, as expected from the diffraction limit. The major advantage of our approach as we have established in this work is that it is orders of magnitude faster than any current localization technique.
We would like to emphasize that this work is meant to be a feasibility test of our method. We plan to try with real LIGO noise and realistic PSD and compare our results with Bayestar and PE in our future work. We also plan to work with higher numbers of sectors and ensure that the model has enough training examples for higher sector numbers by having a larger and more uniform training dataset. Another potential improvement of the localization can be done by increasing the data sampling rate. We will explore the performance of our network for binary neutron star mergers and lower mass black holes in the future.  
 
\begin{acknowledgments}
We wish to thank the support of OzGrav and the University of Western Australia for funding this research. This research was undertaken with the support of computational resources from the Pople high-performance computing cluster of the Faculty of Science at the University of Western Australia. Chayan Chatterjee would like to thank Timothy D. Gebhard and Niki Kilbertus for their immense help and discussions with the GW sample generation code.
\end{acknowledgments}

%\end{thebibliography}

%\appendix*
%\input{sections/appendix1.tex}


\begin{thebibliography}{92}

\bibitem{PhysRevLett.116.131103}
B.~P. Abbott, R.~Abbott, T.~D. Abbott, M.~R. Abernathy, F.~Acernese, K.~Ackley,
  C.~Adams, T.~Adams, P.~Addesso, R.~X. Adhikari, et~al.
\newblock {\em Phys. Rev. Lett.}, 116:131103, Mar 2016.

\bibitem{LIGO}
J~Aasi et~al.
\newblock Advanced {LIGO}.
\newblock {\em Classical and Quantum Gravity}, 32(7):074001, Mar 2015.

\bibitem{virgo}
F.~Acernese et~al.
\newblock {\em Classical and Quantum Gravity}, 32:024001, 2015.

\bibitem{bp1}
B.~P. Abbott, R.~Abbott, T.~D. Abbott, M.~R. Abernathy, F.~Acernese, K.~Ackley,
  C.~Adams, T.~Adams, P.~Addesso, R.~X. Adhikari, et~al.
\newblock Observation of gravitational waves from a binary black hole merger.
\newblock {\em Phys. Rev. Lett.}, 116:061102, Feb 2016.

\bibitem{bp2}
B.~P. Abbott, R.~Abbott, T.~D. Abbott, M.~R. Abernathy, F.~Acernese, K.~Ackley,
  C.~Adams, T.~Adams, P.~Addesso, R.~X. Adhikari, et~al.
\newblock Gw151226: Observation of gravitational waves from a 22-solar-mass
  binary black hole coalescence.
\newblock {\em Phys. Rev. Lett.}, 116:241103, Jun 2016.

\bibitem{bp3}
B.~P. Abbott, R.~Abbott, T.~D. Abbott, M.~R. Abernathy, F.~Acernese, K.~Ackley,
  C.~Adams, T.~Adams, P.~Addesso, R.~X. Adhikari, et~al.
\newblock Gw170104: Observation of a 50-solar-mass binary black hole
  coalescence at redshift 0.2.
\newblock {\em Phys. Rev. Lett.}, 118:221101, Jun 2017.

\bibitem{bp4}
B.~P. Abbott, R.~Abbott, T.~D. Abbott, M.~R. Abernathy, F.~Acernese, K.~Ackley,
  C.~Adams, T.~Adams, P.~Addesso, R.~X. Adhikari, et~al.
\newblock Gw170814: A three-detector observation of gravitational waves from a
  binary black hole coalescence.
\newblock {\em Phys. Rev. Lett.}, 119:141101, Oct 2017.

\bibitem{bp5}
B.~P. Abbott, R.~Abbott, T.~D. Abbott, M.~R. Abernathy, F.~Acernese, K.~Ackley,
  C.~Adams, T.~Adams, P.~Addesso, R.~X. Adhikari, et~al.
\newblock Gw170817: Observation of gravitational waves from a binary neutron
  star inspiral.
\newblock {\em Phys. Rev. Lett.}, 119:161101, Oct 2017.

\bibitem{bp6}
B.~P. Abbott, R.~Abbott, T.~D. Abbott, M.~R. Abernathy, F.~Acernese, K.~Ackley,
  C.~Adams, T.~Adams, P.~Addesso, R.~X. Adhikari, et~al.
\newblock Multi-messenger observations of a binary neutron star merger.
\newblock {\em The Astrophysical Journal}, 848(2):L12, oct 2017.

\bibitem{bp7}
D.~A. Coulter, R.~J. Foley, C.~D. Kilpatrick, M.~R. Drout, A.~L. Piro, B.~J.
  Shappee, M.~R. Siebert, J.~D. Simon, N.~Ulloa, D.~Kasen, B.~F. Madore,
  A.~Murguia-Berthier, Y.-C. Pan, J.~X. Prochaska, E.~Ramirez-Ruiz, A.~Rest,
  and C.~Rojas-Bravo.
\newblock Swope supernova survey 2017a (sss17a), the optical counterpart to a
  gravitational wave source.
\newblock {\em Science}, 358(6370):1556--1558, 2017.

\bibitem{astro_1}
Dark Energy~Survey Collaboration.
\newblock {\em MNRAS}, 460:1270, 2016.

\bibitem{astro_2}
A.~A. Abdo, M.~Ajello, A.~Allafort, L.~Baldini, J.~Ballet, G.~Barbiellini,
  M.~G. Baring, D.~Bastieri, A.~Belfiore, et~al.
\newblock {\em ApJ S}, 208:17, 2013.

\bibitem{astro_3}
J.~A. Tyson.
\newblock {\em Survey and Other Telescope Technologies and Discoveries}, volume
  4863.
\newblock Proceedings of SPIE, 2002.

\bibitem{em_1}
C.~D. Ott.
\newblock {\em Classical and Quantum Gravity}, 26:063001, 2009.

\bibitem{em_2}
R~Haas, C.~D. Ott, et~al.
\newblock {\em Phys. Rev. D}, 93:124062, 2016.

\bibitem{em_3}
P.~M{\"o}sta, B.~C. Mundim, J.~A. Faber, et~al.
\newblock {\em Classical and Quantum Gravity}, 31:015005, 2014.

\bibitem{em_4}
E.~Abdikamalov, S.~Gossan, A.~M. DeMaio, C.~D. Ott, et~al.
\newblock {\em Phys. Rev. D}, 90:044001, 2014.

\bibitem{em_5}
L.~E. Kidder, S.~E. Field, F.~Foucart, E.~Schnetter, S.~A. Teukolsky, A.~Bohn,
  N.~Deppe, P.~Diener, F.~Hébert, J.~Lippuner, J.~Miller, C.~D. Ott, M.~A.
  Scheel, and T.~Vincent.
\newblock {\em Journal of Computational Physics}, 335:84, 2017.

\bibitem{em_6}
S.~Nissanke, M.~Kasliwal, and A.~Georgieva.
\newblock {\em Astrophys. J.}, 767:124, 2013.

\bibitem{mf1}
Samantha~A Usman, Alexander~H Nitz, Ian~W Harry, Christopher~M Biwer, Duncan~A
  Brown, Miriam Cabero, Collin~D Capano, Tito~Dal Canton, Thomas Dent, Stephen
  Fairhurst, Marcel~S Kehl, Drew Keppel, Badri Krishnan, Amber Lenon, Andrew
  Lundgren, Alex~B Nielsen, Larne~P Pekowsky, Harald~P Pfeiffer, Peter~R
  Saulson, Matthew West, and Joshua~L Willis.
\newblock The {PyCBC} search for gravitational waves from compact binary
  coalescence.
\newblock {\em Classical and Quantum Gravity}, 33(21):215004, oct 2016.

\bibitem{mf2}
Kipp Cannon, Romain Cariou, Adrian Chapman, Mireia Crispin-Ortuzar, Nickolas
  Fotopoulos, Melissa Frei, Chad Hanna, Erin Kara, Drew Keppel, Laura Liao,
  Stephen Privitera, Antony Searle, Leo Singer, and Alan Weinstein.
\newblock {\em The Astrophysical Journal}, 748(2):136, mar 2012.

\bibitem{mf3}
Tito Dal~Canton, Alexander~H. Nitz, Andrew~P. Lundgren, Alex~B. Nielsen,
  Duncan~A. Brown, Thomas Dent, Ian~W. Harry, Badri Krishnan, Andrew~J. Miller,
  Karl Wette, Karsten Wiesner, and Joshua~L. Willis.
\newblock Implementing a search for aligned-spin neutron star-black hole
  systems with advanced ground based gravitational wave detectors.
\newblock {\em Phys. Rev. D}, 90:082004, Oct 2014.

\bibitem{mf4}
Duncan~A. Brown, Ian Harry, Andrew Lundgren, and Alexander~H. Nitz.
\newblock Detecting binary neutron star systems with spin in advanced
  gravitational-wave detectors.
\newblock {\em Phys. Rev. D}, 86:084017, Oct 2012.

\bibitem{mf5}
Tito~Dal Canton and Ian~W. Harry.
\newblock Designing a template bank to observe compact binary coalescences in
  advanced ligo's second observing run.
\newblock 2017.

\bibitem{mf6}
I.~W. Harry, B.~Allen, and B.~S. Sathyaprakash.
\newblock Stochastic template placement algorithm for gravitational wave data
  analysis.
\newblock {\em Phys. Rev. D}, 80:104014, Nov 2009.

\bibitem{mf7}
P.~Ajith, N.~Fotopoulos, S.~Privitera, A.~Neunzert, N.~Mazumder, and A.~J.
  Weinstein.
\newblock Effectual template bank for the detection of gravitational waves from
  inspiralling compact binaries with generic spins.
\newblock {\em Phys. Rev. D}, 89:084041, Apr 2014.

\bibitem{mf8}
B.~S. Sathyaprakash and S.~V. Dhurandhar.
\newblock Choice of filters for the detection of gravitational waves from
  coalescing binaries.
\newblock {\em Phys. Rev. D}, 44:3819--3834, Dec 1991.

\bibitem{pn1}
Luc Blanchet.
\newblock Gravitational radiation from post-newtonian sources and inspiralling
  compact binaries.
\newblock {\em Living Reviews in Relativity}, 17(1):2, Feb 2014.

\bibitem{pn2}
K.~G. Arun, Alessandra Buonanno, Guillaume Faye, and Evan Ochsner.
\newblock {\em Phys. Rev. D}, 84:049901, Aug 2011.

\bibitem{pn3}
Alessandra Buonanno, Bala~R. Iyer, Evan Ochsner, Yi~Pan, and B.~S.
  Sathyaprakash.
\newblock {\em Phys. Rev. D}, 80:084043, Oct 2009.

\bibitem{pn4}
Chandra~Kant Mishra, Aditya Kela, K.~G. Arun, and Guillaume Faye.
\newblock {\em Phys. Rev. D}, 93:084054, Apr 2016.

\bibitem{eob}
A.~Buonanno and T.~Damour.
\newblock Effective one-body approach to general relativistic two-body
  dynamics.
\newblock {\em Phys. Rev. D}, 59:084006, Mar 1999.

\bibitem{NumRel}
Frans Pretorius.
\newblock Evolution of binary black-hole spacetimes.
\newblock {\em Phys. Rev. Lett.}, 95:121101, Sep 2005.

\bibitem{pe_fast}
Rory Smith, Scott~E. Field, Kent Blackburn, Carl-Johan Haster, Michael
  P\"urrer, Vivien Raymond, and Patricia Schmidt.
\newblock Fast and accurate inference on gravitational waves from precessing
  compact binaries.
\newblock {\em Phys. Rev. D}, 94:044031, Aug 2016.

\bibitem{deeplearning1}
Ian~J. Goodfellow, Jean Pouget-Abadie, Mehdi Mirza, Bing Xu, David
  Warde-Farley, Sherjil Ozair, Aaron Courville, and Yoshua Bengio.
\newblock Generative adversarial networks.
\newblock 2014.

\bibitem{deeplearning2}
Karen Simonyan and Andrew Zisserman.
\newblock Very deep convolutional networks for large-scale image recognition.

\bibitem{deeplearning4}
Liang-Chieh Chen, George Papandreou, Iasonas Kokkinos, Kevin Murphy, and
  Alan~L. Yuille.
\newblock Semantic image segmentation with deep convolutional nets and fully
  connected crfs.
\newblock 2014.

\bibitem{deeplearning5}
Matthew~D Zeiler and Rob Fergus.
\newblock Visualizing and understanding convolutional networks.
\newblock 2013.

\bibitem{deeplearning6}
Christian Szegedy, Wei Liu, Yangqing Jia, Pierre Sermanet, Scott Reed, Dragomir
  Anguelov, Dumitru Erhan, Vincent Vanhoucke, and Andrew Rabinovich.
\newblock Going deeper with convolutions.
\newblock 2014.

\bibitem{O3localization}
B.~P. Abbott, R.~Abbott, T.~D. Abbott, M.~R. Abernathy, F.~Acernese, K.~Ackley,
  C.~Adams, T.~Adams, P.~Addesso, R.~X. Adhikari, V.~B. Adya, C.~Affeldt,
  M.~Agathos, K.~Agatsuma, N.~Aggarwal, O.~D. Aguiar, L.~Aiello, A.~Ain,
  P.~Ajith, T.~Akutsu, B.~Allen, A.~Allocca, P.~A. Altin, A.~Ananyeva, S.~B.
  Anderson, W.~G. Anderson, M.~Ando, S.~Appert, K.~Arai, A.~Araya, M.~C. Araya,
  J.~S. Areeda, N.~Arnaud, K.~G. Arun, H.~Asada, S.~Ascenzi, G.~Ashton, Y.~Aso,
  M.~Ast, S.~M. Aston, P.~Astone, S.~Atsuta, P.~Aufmuth, C.~Aulbert,
  A.~Avila-Alvarez, K.~Awai, S.~Babak, P.~Bacon, M.~K.~M. Bader, L.~Baiotti,
  P.~T. Baker, F.~Baldaccini, G.~Ballardin, S.~W. Ballmer, J.~C. Barayoga,
  S.~E. Barclay, B.~C. Barish, D.~Barker, F.~Barone, B.~Barr, L.~Barsotti,
  M.~Barsuglia, D.~Barta, J.~Bartlett, M.~A. Barton, I.~Bartos, R.~Bassiri,
  A.~Basti, J.~C. Batch, C.~Baune, V.~Bavigadda, M.~Bazzan, B.~B{\'e}csy,
  C.~Beer, M.~Bejger, I.~Belahcene, M.~Belgin, A.~S. Bell, B.~K. Berger,
  G.~Bergmann, C.~P.~L. Berry, D.~Bersanetti, A.~Bertolini, J.~Betzwieser,
  S.~Bhagwat, R.~Bhandare, I.~A. Bilenko, G.~Billingsley, C.~R. Billman,
  J.~Birch, R.~Birney, O.~Birnholtz, S.~Biscans, A.~Bisht, M.~Bitossi,
  C.~Biwer, M.~A. Bizouard, J.~K. Blackburn, J.~Blackman, C.~D. Blair, D.~G.
  Blair, R.~M. Blair, S.~Bloemen, O.~Bock, M.~Boer, G.~Bogaert, A.~Bohe,
  F.~Bondu, R.~Bonnand, B.~A. Boom, R.~Bork, V.~Boschi, S.~Bose, Y.~Bouffanais,
  A.~Bozzi, C.~Bradaschia, P.~R. Brady, V.~B. Braginsky, M.~Branchesi, J.~E.
  Brau, T.~Briant, A.~Brillet, M.~Brinkmann, V.~Brisson, P.~Brockill, J.~E.
  Broida, A.~F. Brooks, D.~A. Brown, D.~D. Brown, N.~M. Brown, S.~Brunett,
  C.~C. Buchanan, A.~Buikema, T.~Bulik, H.~J. Bulten, A.~Buonanno, D.~Buskulic,
  C.~Buy, R.~L. Byer, M.~Cabero, L.~Cadonati, G.~Cagnoli, C.~Cahillane,
  J.~Calder{\'o}n Bustillo, T.~A. Callister, E.~Calloni, J.~B. Camp, K.~C.
  Cannon, H.~Cao, J.~Cao, C.~D. Capano, E.~Capocasa, F.~Carbognani, S.~Caride,
  J.~Casanueva Diaz, C.~Casentini, S.~Caudill, M.~Cavagli{\`a}, F.~Cavalier,
  R.~Cavalieri, G.~Cella, C.~B. Cepeda, L.~Cerboni Baiardi, G.~Cerretani,
  E.~Cesarini, S.~J. Chamberlin, M.~Chan, S.~Chao, P.~Charlton,
  E.~Chassande-Mottin, B.~D. Cheeseboro, H.~Y. Chen, Y.~Chen, H.-P. Cheng,
  A.~Chincarini, A.~Chiummo, T.~Chmiel, H.~S. Cho, M.~Cho, J.~H. Chow,
  N.~Christensen, Q.~Chu, A.~J.~K. Chua, S.~Chua, S.~Chung, G.~Ciani, F.~Clara,
  J.~A. Clark, F.~Cleva, C.~Cocchieri, E.~Coccia, P.-F. Cohadon, A.~Colla,
  C.~G. Collette, L.~Cominsky, M.~Constancio, L.~Conti, S.~J. Cooper, T.~R.
  Corbitt, N.~Cornish, A.~Corsi, S.~Cortese, C.~A. Costa, M.~W. Coughlin, S.~B.
  Coughlin, J.-P. Coulon, S.~T. Countryman, P.~Couvares, P.~B. Covas, E.~E.
  Cowan, D.~M. Coward, M.~J. Cowart, D.~C. Coyne, R.~Coyne, J.~D.~E. Creighton,
  T.~D. Creighton, J.~Cripe, S.~G. Crowder, T.~J. Cullen, A.~Cumming,
  L.~Cunningham, E.~Cuoco, T.~Dal Canton, S.~L. Danilishin, S.~D'Antonio,
  K.~Danzmann, A.~Dasgupta, C.~F. Da~Silva Costa, V.~Dattilo, I.~Dave,
  M.~Davier, G.~S. Davies, D.~Davis, E.~J. Daw, B.~Day, R.~Day, S.~De,
  D.~DeBra, G.~Debreczeni, J.~Degallaix, M.~De~Laurentis, S.~Del{\'e}glise,
  W.~Del~Pozzo, T.~Denker, T.~Dent, V.~Dergachev, R.~De~Rosa, R.~T. DeRosa,
  R.~DeSalvo, R.~C. Devine, S.~Dhurandhar, M.~C. D{\'i}az, L.~Di Fiore, M.~Di
  Giovanni, T.~Di Girolamo, A.~Di Lieto, S.~Di Pace, I.~Di Palma, A.~Di
  Virgilio, Z.~Doctor, K.~Doi, V.~Dolique, F.~Donovan, K.~L. Dooley,
  S.~Doravari, I.~Dorrington, R.~Douglas, M.~Dovale {\'A}lvarez, T.~P. Downes,
  M.~Drago, R.~W.~P. Drever, J.~C. Driggers, Z.~Du, M.~Ducrot, S.~E. Dwyer,
  K.~Eda, T.~B. Edo, M.~C. Edwards, A.~Effler, H.-B. Eggenstein, P.~Ehrens,
  J.~Eichholz, S.~S. Eikenberry, R.~A. Eisenstein, R.~C. Essick, Z.~Etienne,
  T.~Etzel, M.~Evans, T.~M. Evans, R.~Everett, M.~Factourovich, V.~Fafone,
  H.~Fair, S.~Fairhurst, X.~Fan, S.~Farinon, B.~Farr, W.~M. Farr, E.~J.
  Fauchon-Jones, and M.~Favata.
\newblock Prospects for observing and localizing gravitational-wave transients
  with advanced ligo, advanced virgo and kagra.
\newblock {\em Living Reviews in Relativity}, 21(1):3, Apr 2018.

\bibitem{ANN}
Kevin~L. Priddy and Paul~E. Keller.
\newblock {\em Artificial Neural Networks: An Introduction (SPIE Tutorial Texts
  in Optical Engineering, Vol. TT68)}.
\newblock SPIE- International Society for Optical Engineering, 2005.

\bibitem{ComputationallyIntensive}
Ian Harry, Stephen Privitera, Alejandro Boh\'e, and Alessandra Buonanno.
\newblock Searching for gravitational waves from compact binaries with
  precessing spins.
\newblock {\em Phys. Rev. D}, 94:024012, Jul 2016.

\bibitem{image1}
Richard Zhang, Phillip Isola, and Alexei~A. Efros.
\newblock Colorful image colorization.
\newblock 2016.

\bibitem{image2}
Andrej Karpathy and Li~Fei-Fei.
\newblock Deep visual-semantic alignments for generating image descriptions.
\newblock 2014.

\bibitem{detection1}
Antonis Mytidis, Athanasios~Aris Panagopoulos, Orestis~P. Panagopoulos, Andrew
  Miller, and Bernard Whiting.
\newblock {\em Phys. Rev. D}, 99:024024, Jan 2019.

\bibitem{detection2}
Daniel George, Hongyu Shen, and E.~A. Huerta.
\newblock {\em Phys. Rev. D}, 97:101501, May 2018.

\bibitem{detection4}
Xiangru Li, Woliang Yu, and Xilong Fan.
\newblock 2017.

\bibitem{detection5}
Daniel George and E.~A. Huerta.
\newblock 2017.

\bibitem{detection6}
Kyungmin Kim, Ian~W Harry, Kari~A Hodge, Young-Min Kim, Chang-Hwan Lee,
  Hyun~Kyu Lee, John~J Oh, Sang~Hoon Oh, and Edwin~J Son.
\newblock {\em Classical and Quantum Gravity}, 32(24):245002, nov 2015.

\bibitem{detection7}
Daniel George and E.~A. Huerta.
\newblock Deep neural networks to enable real-time multimessenger astrophysics.
\newblock {\em Phys. Rev. D}, 97:044039, Feb 2018.

\bibitem{detection8}
Fan, XiLong and Li, Jin and Li, Xin and Zhong, YuanHong and Cao, JunWei
\newblock Applying deep neural networks to the detection and space parameter estimation of compact binary coalescence with a network of gravitational wave detectors.
\newblock {\em Science China Physics, Mechanics {\&} Astronomy}, 10.1007/s11433-018-9321-7.


\bibitem{glitch1}
B.~Sánchez, M.J.~Domínguez R., M.~Lares, M.~Beroiz, J.B. Cabral, S.~Gurovich,
  C.~Quiñones, R.~Artola, C.~Colazo, M.~Schneiter, C.~Girardini, M.~Tornatore,
  J.L.~Nilo Castellón, D.~García Lambas, and M.C. Díaz.
\newblock Machine learning on difference image analysis: A comparison of
  methods for transient detection.
\newblock {\em Astronomy and Computing}, 28:100284, 2019.

\bibitem{glitch2}
Daniel George, Hongyu Shen, and E.~A. Huerta.
\newblock Classification and unsupervised clustering of ligo data with deep
  transfer learning.
\newblock {\em Phys. Rev. D}, 97:101501, May 2018.

\bibitem{glitch3}
Sara Bahaadini, Neda Rohani, Scott Coughlin, Michael Zevin, Vicky Kalogera, and
  Aggelos~K Katsaggelos.
\newblock Deep multi-view models for glitch classification.
\newblock 2017.

\bibitem{glitch4}
Jade Powell, Daniele Trifir{\`{o}}, Elena Cuoco, Ik~Siong Heng, and Marco
  Cavagli{\`{a}}.
\newblock Classification methods for noise transients in advanced
  gravitational-wave detectors.
\newblock {\em Classical and Quantum Gravity}, 32(21):215012, oct 2015.

\bibitem{glitch5}
Miquel Llorens-Monteagudo, Alejandro Torres-Forn{\'{e}}, Jos{\'{e}}~A Font, and
  Antonio Marquina.
\newblock Classification of gravitational-wave glitches via dictionary
  learning.
\newblock {\em Classical and Quantum Gravity}, 36(7):075005, mar 2019.

\bibitem{glitch6}
Massimiliano Razzano and Elena Cuoco.
\newblock Image-based deep learning for classification of noise transients in gravitational wave detectors
\newblock {\em Classical and Quantum Gravity}, 10.1088/1361-6382/aab793.


\bibitem{denoising1}
Hongyu Shen, Daniel George, E.~A. Huerta, and Zhizhen Zhao.
\newblock Denoising gravitational waves with enhanced deep recurrent denoising
  auto-encoders.

\bibitem{denoising2}
Marco Cavaglia, Kai Staats, and Teerth Gill.
\newblock Finding the origin of noise transients in ligo data with machine
  learning.
\newblock 2018.

\bibitem{denoising3}
Alejandro Torres-Forn\'e, Elena Cuoco, Antonio Marquina, Jos\'e~A. Font, and
  Jos\'e~M. Ib\'a\~nez.
\newblock Total-variation methods for gravitational-wave denoising: Performance
  tests on advanced ligo data.
\newblock {\em Phys. Rev. D}, 98:084013, Oct 2018.

\bibitem{denoising4}
Hongyu Shen, Daniel George, E.~A. Huerta, and Zhizhen Zhao.

\bibitem{multipole}
Adam Rebei, E.~A. Huerta, Sibo Wang, Sarah Habib, Roland Haas, Daniel Johnson,
  and Daniel George.
\newblock Fusing numerical relativity and deep learning to detect higher-order
  multipole waveforms from eccentric binary black hole mergers.
\newblock 2018.

\bibitem{EMBlackHole1}
D.~V. Bisikalo, A.~G. Zhilkin, and E.~P. Kurbatov.
\newblock Possible electromagnetic manifestations of merging black holes.
\newblock {\em Astronomy Reports}, 63(1):1--14, Jan 2019.

\bibitem{EMBlackHole2}
Agnieszka Janiuk, M.~Bejger, S.~Charzyński, and P.~Sukova.
\newblock On the possible gamma-ray burst–gravitational wave association in
  gw150914.
\newblock {\em New Astronomy}, 51:7 -- 14, 2017.

\bibitem{EMBlackHole3}
Abraham Loeb.
\newblock {ELECTROMAGNETIC} {COUNTERPARTS} {TO} {BLACK} {HOLE} {MERGERS}
  {DETECTED} {BY} {LIGO}.
\newblock {\em The Astrophysical Journal}, 819(2):L21, mar 2016.

\bibitem{EMBlackHole4}
S.~E. de~Mink and A.~King.
\newblock Electromagnetic signals following stellar-mass black hole mergers.
\newblock {\em The Astrophysical Journal}, 839(1):L7, apr 2017.

\bibitem{EMBlackHole5}
Rosalba Perna, Davide Lazzati, and Bruno Giacomazzo.
\newblock {SHORT} {GAMMA}-{RAY} {BURSTS} {FROM} {THE} {MERGER} {OF} {TWO}
  {BLACK} {HOLES}.
\newblock {\em The Astrophysical Journal}, 821(1):L18, apr 2016.

\bibitem{localise1}
Stephen Fairhurst.
\newblock Triangulation of gravitational wave sources with a network of
  detectors.
\newblock {\em New Journal of Physics}, 13(6):069602, jun 2011.

\bibitem{localise2}
Yekta G\"ursel and Massimo Tinto.
\newblock Near optimal solution to the inverse problem for gravitational-wave
  bursts.
\newblock {\em Phys. Rev. D}, 40:3884--3938, Dec 1989.

\bibitem{localise3}
Antony~C Searle, Patrick~J Sutton, Massimo Tinto, and Graham Woan.
\newblock {\em Classical and Quantum Gravity}, 25(11):114038, may 2008.

\bibitem{localise4}
Antony~C Searle, Patrick~J Sutton, and Massimo Tinto.
\newblock {\em Classical and Quantum Gravity}, 26(15):155017, jul 2009.

\bibitem{localise5}
Fabien Cavalier, Matteo Barsuglia, Marie-Anne Bizouard, Violette Brisson,
  Andr\'e-Claude Clapson, Michel Davier, Patrice Hello, Stephane Kreckelbergh,
  Nicolas Leroy, and Monica Varvella.
\newblock {\em Phys. Rev. D}, 74:082004, Oct 2006.

\bibitem{localise6}
L~Wen, X~Fan, and Y~Chen.
\newblock {\em Journal of Physics: Conference Series}, 122:012038, jul 2008.

\bibitem{localise7}
F~Beauville, M-A Bizouard, L~Blackburn, L~Bosi, P~Brady, L~Brocco, D~Brown,
  D~Buskulic, F~Cavalier, S~Chatterji, N~Christensen, A-C Clapson, S~Fairhurst,
  D~Grosjean, G~Guidi, P~Hello, E~Katsavounidis, M~Knight, A~Lazzarini,
  N~Leroy, F~Marion, B~Mours, F~Ricci, A~Vicer{\'{e}}, M~Zanolin, and The joint
  LIGO/Virgo~working group.
\newblock {\em Journal of Physics: Conference Series}, 32:212--222, mar 2006.

\bibitem{localise8}
S~Birindelli et~al.
\newblock {\em Coherent algorithm for reconstructing the location of a
  coalescing binary using a system of three gravitational wave
  interferometers}.
\newblock {PhD} dissertation, University of Pisa, 2008.

\bibitem{localise9}
J.~Markowitz, M.~Zanolin, L.~Cadonati, and E.~Katsavounidis.
\newblock {\em Phys. Rev. D}, 78:122003, Dec 2008.

\bibitem{convwave}
Timothy~D. Gebhard, Niki Kilbertus, Ian Harry, and Bernhard Schölkopf.
\newblock Convolutional neural networks: a magic bullet for gravitational-wave
  detection?
\newblock 2019.

\bibitem{SEOBNRv4}
Alejandro Boh\'e, Lijing Shao, Andrea Taracchini, Alessandra Buonanno,
  Stanislav Babak, Ian~W. Harry, Ian Hinder, Serguei Ossokine, Michael
  P\"urrer, Vivien Raymond, Tony Chu, Heather Fong, Prayush Kumar, Harald~P.
  Pfeiffer, Michael Boyle, Daniel~A. Hemberger, Lawrence~E. Kidder, Geoffrey
  Lovelace, Mark~A. Scheel, and B\'ela Szil\'agyi.
\newblock {\em Phys. Rev. D}, 95:044028, Feb 2017.

\bibitem{antennabook1}
J.~D.~E. Creighton and W.~G. Anderson.
\newblock chapter Gravitational Wave Detectors, pages 241--244.
\newblock Wiley-VCH, Weinheim, 2011.

\bibitem{antennabook2}
M.~Maggiore.
\newblock chapter Data Analysis Techniques, pages 339--343.
\newblock Oxford University Press, New York, 2008.

\bibitem{antennabook3}
K.~S. Thorne.
\newblock pages 330--458.
\newblock Cambridge University Press, Cambridge, 2008.

\bibitem{alex_nitz_2019_2581446}
Alex Nitz, Ian Harry, Duncan Brown, Christopher~M. Biwer, Josh Willis, Tito~Dal
  Canton, Larne Pekowsky, Collin Capano, Thomas Dent, Andrew~R. Williamson,
  Soumi De, Miriam Cabero, Bernd Machenschalk, Prayush Kumar, Steven Reyes,
  Thomas Massinger, Duncan Macleod, Amber Lenon, Stephen Fairhurst, Sebastian
  Khan, GarethDaviesGW, Alex Nielsen, shasvath, dfinstad, Francesco Pannarale,
  Leo Singer, Márton Tápai, Hunter Gabbard, Peter Couvares, and
  Lorena~Magaña Zertuche.
\newblock gwastro/pycbc: Pycbc release v1.13.5, March 2019.

\bibitem{Pandas}
Eric Jones, Travis Oliphant, Pearu Peterson, et~al.
\newblock {SciPy}: Open source scientific tools for {Python}, 2001--.
\newblock [Online; accessed <today>].

\bibitem{Keras}
Fran\c{c}ois Chollet et~al.
\newblock Keras.
\newblock \url{https://github.com/fchollet/keras}, 2015.

\bibitem{Tensorflow}
Mart\'{\i}n Abadi, Ashish Agarwal, Paul Barham, Eugene Brevdo, Zhifeng Chen,
  Craig Citro, Greg~S. Corrado, Andy Davis, Jeffrey Dean, Matthieu Devin,
  Sanjay Ghemawat, Ian Goodfellow, Andrew Harp, Geoffrey Irving, Michael Isard,
  Yangqing Jia, Rafal Jozefowicz, Lukasz Kaiser, Manjunath Kudlur, Josh
  Levenberg, Dan Man\'{e}, Rajat Monga, Sherry Moore, Derek Murray, Chris Olah,
  Mike Schuster, Jonathon Shlens, Benoit Steiner, Ilya Sutskever, Kunal Talwar,
  Paul Tucker, Vincent Vanhoucke, Vijay Vasudevan, Fernanda Vi\'{e}gas, Oriol
  Vinyals, Pete Warden, Martin Wattenberg, Martin Wicke, Yuan Yu, and Xiaoqiang
  Zheng.
\newblock {TensorFlow}: Large-scale machine learning on heterogeneous systems,
  2015.
\newblock Software available from tensorflow.org.

\bibitem{adam}
Diederik~P. Kingma and Jimmy Ba.
\newblock Adam: A method for stochastic optimization.
\newblock 2014.

\bibitem{curri1}
Yoshua Bengio, J{\'e}r\^{o}me Louradour, Ronan Collobert, and Jason Weston.
\newblock Curriculum learning.
\newblock In {\em Proceedings of the 26th Annual International Conference on
  Machine Learning}, ICML '09, pages 41--48, New York, NY, USA, 2009. ACM.

\bibitem{detection3}
Daniel George and E.A. Huerta.
\newblock {\em Physics Letters B}, 778:64 -- 70, 2018.

\bibitem{curri2}
Wei Wei and E.~A. Huerta.
\newblock Gravitational wave denoising of binary black hole mergers with deep
  learning.
\newblock 2019.

\bibitem{curri3}
Hongyu Shen, E.~A. Huerta, and Zhizhen Zhao.
\newblock Deep learning at scale for gravitational wave parameter estimation of
  binary black hole mergers.

\bibitem{hilbert1}
L.~Cohen.
\newblock {\em Time-frequency Analysis}.
\newblock Electrical engineering signal processing. Prentice Hall PTR, 1995.

\bibitem{hilbert2}
A.V. Oppenheim.
\newblock {\em Discrete-Time Signal Processing}.
\newblock Pearson education signal processing series. Pearson Education, 1999.

\bibitem{LeoSinger}
Leo~P. Singer and Larry~R. Price.
\newblock Rapid bayesian position reconstruction for gravitational-wave
  transients.
\newblock {\em Phys. Rev. D}, 93:024013, Jan 2016.

\bibitem{Wen&Chen}
Linqing Wen and Yanbei Chen.
\newblock Geometrical expression for the angular resolution of a network of
  gravitational-wave detectors.
\newblock {\em Phys. Rev. D}, 81:082001, Apr 2010.

\bibitem{GW170818}
B.~P. Abbott, R.~Abbott, T.~D. Abbott, M.~R. Abernathy, F.~Acernese, K.~Ackley,
  C.~Adams, T.~Adams, P.~Addesso, R.~X. Adhikari, et~al.
\newblock GWTC-1: A Gravitational-Wave Transient Catalog of Compact Binary Mergers Observed by LIGO and Virgo during the First and Second Observing Runs
\newblock {\em Phys. Rev. X}, 10.1103/PhysRevX.9.031040


\end{thebibliography}
\end{document}